\documentclass[fleqn,usenatbib]{mnras}

\usepackage{newtxtext,newtxmath}
\usepackage[dvipsnames]{xcolor}
\usepackage{ulem}
\usepackage[T1]{fontenc}

\DeclareRobustCommand{\VAN}[3]{#2}
\let\VANthebibliography\thebibliography
\def\thebibliography{\DeclareRobustCommand{\VAN}[3]{##3}\VANthebibliography}

\usepackage{graphicx}	
\usepackage{amsmath}	
\usepackage{comment}
\usepackage{graphicx}
\usepackage{dcolumn}
\usepackage{bm}
\usepackage{stfloats}
\newcommand{\HI}{\rm H{\sc i}~}
\newcommand{\HII}{\rm H~{\sc ii }}

\newcommand{\MSUN}{{\rm M}_{\odot}}

\newcommand{\XHII}{x_{\rm HII}}
\newcommand{\TS}{T_{s}}

\newcommand{\TK}{T_{ k}}

\newcommand{\LYA}{\rm {Ly{\alpha}}}
\newcommand{\OmegaB}{\Omega_{\rm b}}
\newcommand{\Omegam}{\Omega_{\rm m}}

\newcommand{\DTB}{\delta T_{\rm b}}
\newcommand{\MHMIN}{M_{\rm h,min}}
\newcommand{\FX}{f_{\rm x}}

\title[\HI signal optical depth]{The Effect of Large Optical Depths on the Non-Gaussian 21-cm signal from Cosmic Dawn}

\author[Nasreen et al.]{
Iffat Nasreen$^{1}$, 
Kanan K. Datta$^{1}$\thanks{E-mail: kanankdatta.physics@jadavpuruniversity.in}, 
Abinash Kumar Shaw$^{2}$, 
Leon Noble$^{3}$, 
Raghunath Ghara$^{4}$,
Sk. Saiyad Ali$^{1}$, 
\newauthor 
Arnab Mishra$^{1}$, 
Mohd Kamran$^{5, 6, 7}$
and Suman Majumdar$^{3}$\\
$^{1}$Relativity and Cosmology Research Centre (RCRC), Department of Physics, Jadavpur University, Kolkata 700032, India\\
$^{2}$Department of Computer Science, University of Nevada, Las Vegas, Nevada 89154, USA\\
$^{3}$Department of Astronomy, Astrophysics and Space Engineering, Indian Institute of Technology Indore, Indore 453552, India\\
$^{4}$Department of Physical Sciences, Indian Institute of Science Education and Research Kolkata, Mohanpur, WB 741 246, India\\
$^{5}$INAF – Astronomical Observatory of Trieste, Via G. B. Tiepolo 11, 34143 Trieste, Italy\\
$^{6}$IFPU – Institute for Fundamental Physics of the Universe, Via Beirut 2, 34151 Trieste, Italy\\
$^{7}$Department of Physics and Astronomy, Uppsala University, Uppsala 75237, Sweden
}

\date{Accepted XXX. Received YYY; in original form ZZZ}

\pubyear{\the\year{}}

\begin{document}
\label{firstpage}
\pagerange{\pageref{firstpage}--\pageref{lastpage}}
\maketitle

\begin{abstract}
During the Cosmic Dawn (CD), the HI 21-cm optical depth ($\tau$ ) in the intergalactic medium can become significantly large. Consequently, the second and higher-order terms of $\tau$ appearing in the Taylor expansion of the HI 21-cm differential brightness temperature ($\delta T_{\rm b}$ ) become important. This  introduces additional non-Gaussianity into the signal. We study the impact of large $\tau$ on statistical quantities of HI 21-cm signal  using a suite of standard numerical simulations that vary X-ray heating efficiency and the minimum halo mass required to host radiation sources. We find that the higher order terms suppress statistical quantities such as skewness, power-spectrum and bispectrum. However, the effect is found to be particularly strong on the non-Gaussian signal. We find that the change in skewness can reach several hundred percent in low X-ray heating scenarios, whereas for moderate and high X-ray heating models the changes are around $\sim 40\%$ and $\sim 60\%$, respectively, for $\MHMIN = 10^{9}\, \MSUN$. This change is around $\sim 75\%$, $25\%$ and $20\%$ for low, moderate and high X-ray heating models, respectively, for $\MHMIN=10^{10}\, \MSUN$.  The change in bispectrum in both the halo cutoff mass scenarios ranges from $\sim 10\%$ to $\sim 300\%$ for low X-ray heating model. However, for moderate and high X-ray heating models the change remains between $\sim 10\%$ to $\sim 200\%$ for both equilateral and squeezed limit triangle configuration. Finally, we find that up to third orders of $\tau$ need
to be retained to accurately model $\delta T_{\rm b}$, especially
for capturing the non-Gaussian features in the \HI 21-cm signal.
\end{abstract}




\section{Introduction}
\label{sec:intro}
Observations of the cosmological \HI 21-cm signal hold great promise for unveiling the details of the Cosmic Dawn (CD) and the Epoch of Reionization (EoR) \citep{Pritchard2012, Bera2023}. Significant efforts are currently underway to study this era through radio observations of the \HI 21-cm line. Ongoing interferometric telescopes such as the Low Frequency Array \citep[LOFAR\footnote{\url{http://www.lofar.org/}};][]{vanHaarlem2013LOFAR:ARray}, the upgraded Giant Metrewave Radio Telescope \citep[uGMRT;][]{ali2008, Ghosh2012, Shaw2023}, the Murchison Widefield Array \citep[MWA\footnote{\url{https://www.mwatelescope.org/telescope/}};][]{tingay13}, the Hydrogen Epoch of Reionization Array \citep[HERA\footnote{\url{https://reionization.org/}};][]{2017PASP..129d5001D} and the New Extension in Nançay Upgrading LOFAR \citep[NenuFAR\footnote{\url{https://nenufar.obs-nancay.fr/en/homepage-en/}};][]{munshi2024} along with the upcoming Square Kilometre Array \citep[SKA\footnote{\url{http://www.skatelescope.org/}};][]{Mellema2013, Koopmans_2015}, have the primary aim of measuring statistical properties of the signal, such as the power spectrum and bispectrum, to probe the early universe. Some upper limits on the 21-cm power spectrum have already been reported using observations with LOFAR \citep{2020MNRAS.493.1662M, lofar2024, 2024MNRAS.534L..30A}, GMRT \citep{paciga13, Pal2021}, PAPER \citep{2018ApJ...868...26C, 2019ApJ...883..133K}, MWA \citep{2019ApJ...884....1B, 2020MNRAS.493.4711T} and HERA \citep{Abdurashidova_2023} 
and are being used to constrain some of the source and  IGM properties during CD-EoR \citep[see e.g.][]{2020MNRAS.498.4178M,  2020MNRAS.493.4728G, 2021MNRAS.503.4551G, 2020arXiv200603203G, 2022ApJ...924...51A, 2025arXiv250500373G}. 
In this context, accurately modelling the statistical properties of the signal is crucial. It not only deepens our understanding of the signal itself but also helps in designing effective observational strategies,  data analysis methods, and ensuring reliable interpretation of any detection.

In this paper, we study the impact of large 21-cm signal optical depth on the spatial fluctuations of the CD 21-cm signal differential brightness temperature ($\DTB$). In particular, we focus on the non-Gaussian features of the CD 21-cm signal, along with its power spectrum. The 21-cm signal optical depth ($\tau$), which is proportional to the \HI density and inversely proportional to the \HI spin temperature, can become significantly large during CD, particularly in regions with high \HI density and low spin temperature. 
As a result, when $\DTB$ is expanded as a Taylor series in $\tau$, not only the linear term (which has been generally taken into account in the literature), but also contributions from the second and higher-order terms in the expansion of $\DTB$ become important \citep[see e.g.,][]{datta2022}. These higher-order terms not only change the spatial fluctuations in $\DTB$, but also introduce additional non-Gaussianity in the signal. Therefore, it is important to investigate how this affects the spatial fluctuations in $\DTB$, as well as the non-Gaussian properties of the signal from CD. 

There have been earlier studies of the non-Gaussianity in the \HI 21-cm signal from CD and reionization eras using statistical measures such as skewness and the skew spectrum \citep{Harker_2009, Patil_2014, 2021MNRAS.506.3717R}. These statistical quantities have been shown to be detectable by the SKA-Low \citep{2023MNRAS.523..640M, 10.1093/mnras/stae593}. 

The bispectrum has been extensively used to quantify the non-Gaussian nature of the \HI 21-cm signal from the CD-EoR \citep{Bharadwaj_2005, Shimabukuro_2016,Majumdar_2018}. 
The bispectrum, a Fourier conjugate of the three-point correlation function, is sensitive to correlations between different Fourier modes. \citet{Majumdar_2020} presented the first comprehensive analysis of the EoR 21-cm bispectrum across all unique $k$-triangle configurations. Several other studies have explored various physical effects on the 21-cm bispectrum, including X-ray heating and $\LYA$ coupling \citep{Watkinson_2018, (Kamran_2021, Kamran_2021, Kamran_2022}, IGM physics \citep{Kamran_2021,Kamran_2022}, redshift-space distortions and light cone effect \citep{(Kamran_2021, Mondal_2021, Gill_2024}, impact of ionizing source models and ionization morphology \citep{Noble_2024,Hutter_2019}, impact of distribution of neutral hydrogen islands at the end stage of reionization \citep{Raste_2024}, impact of different kinds of dark matter models \citep{Saxsena_2020} etc. It has also been demonstrated that the SKA-Low will be sensitive enough to detect the 21-cm bispectrum from the reionization era \citep{Mondal_2021, Tiwari_2022}, and that the bispectrum can place tighter constraints on reionization model parameters compared to the HI 21-cm power spectrum \citep{Tiwari_2022, mahida2025}. 
Efforts have also been made to directly measure the bispectrum of the signal from observational data \citep{Trott_2019, gill25a}.

Given the promising prospects of detecting the CD-EoR 21-cm power spectrum and bispectrum with ongoing and upcoming radio interferometric experiments, and their potential to constrain CD-EoR models, we present a detailed study on the impact of higher-order terms of the \HI optical depth on the spatial fluctuations of $\DTB$, as well as its non-Gaussian features. In our earlier work \citep{datta2022}, we investigate the impact of large \HI 21-cm optical depth on the interpretation of measured global signal. Further, we carried out a limited study on the impact of optical depth corrections on various statistical measures of the $\DTB$ signal, such as its variance, skewness, and power spectrum, at a few redshifts during CD. Here, our analysis focuses specifically on quantifying non-Gaussianity in the signal through skewness and the bispectrum. We use a suite of standard numerical simulations that vary key parameters such as the X-ray heating efficiency and the minimum halo mass required to host radiation sources. These variations result in a broad range of \HI 21-cm optical depths, enabling us to examine the effects across a wide variety of possible scenarios throughout the CD. In addition, we investigate in detail the impact of optical depth corrections on the \HI power spectrum, which remains the primary target of most current experimental efforts.

The outline of the paper is as follows. In Section \ref{sec:2} we will discuss about the fundamentals of the \HI 21-cm signal. We present the basic equations for computing the \HI 21-cm differential brightness temperature, both exactly and approximately as outlined in subsection \ref{sec:2.1}. 
 Subsection \ref{sec:2.2} describes the simulations and models used in this work along with the results on the global evolution of the \HI 21-cm signal. In Section \ref{sec:3} we discuss our results. Subsections \ref{sec:3.1} and \ref{sec:3.2} discusses the optical depth, and its distribution and the impact of large \HI optical depth on the non-Gaussian features of the 21-cm signal, focusing on the distribution of $\DTB$. In subsections \ref{sec:3.3} and \ref{sec:3.4}, we discuss the redshift evolution of the skewness and the impact of large optical depth on it, respectively. Subsections \ref{sec:3.5} and \ref{sec:3.6} examine the redshift evolution of the bispectrum and explores how large optical depths influence it. We also discuss the effect of large optical depth on the \HI 21-cm power spectrum in Section  \ref{sec:4}. Section \ref{sec:5} investigates the effect of including different orders in the optical depth expansion on bispectrum estimation. Finally, Section \ref{sec:6} summarizes the main results of the paper.

In this paper, we use the values of various cosmological parameters for our numerical calculations and simulations: $\Omega_{\rm b}=0.044$, $\Omega_{\rm m}=0.27$, $h=0.7$  and the present Cosmic Microwave Background Radiation (CMBR) temperature of $2.725$ K. These values are consistent with the WMAP results \citep{Hinshaw_2013} and lie within the error bars of the Planck results \citep{2014}.

\section{21-cm signal from CD \& EoR}
\label{sec:2}
In this section, we present the basic equations governing the \HI $21$-cm signal, along with a detailed description of the simulations we use in this study.
\subsection{Basic equations of {\HI} 21-cm signal}
\label{sec:2.1}
The differential brightness temperature corresponding to redshifted \HI $21$-cm signal relative to the CMBR can be written as \citep{rybicki},
\begin{equation}
\DTB({\bf n},z)=\frac{T_S({\bf n},z) -T_\gamma(z)}{1+z}\left[1-\exp\{-\tau({\bf n},z)\}\right],
\label{eq:exact}
\end{equation}
where $T_S({\bf n},z)$ is the \HI spin temperature corresponding to its hyperfine transition and $T_\gamma$ is the CMBR temperature. $\tau({\bf n},z)$ is the optical depth that quantifies the level of absorption or emission of \HI 21-cm line within a medium (here IGM) and can be written as \citep{Pritchard_2008},
\begin{eqnarray}
\tau(\mathbf{n},z) &=& 27\times 10^{-3}\hspace{0.1cm} 
   \frac{x_{\mathrm{HI}}(\mathbf{n},z)}{T_S(\mathbf{n},z)} 
   \left[1 + \delta_{\mathrm{b}}(\mathbf{n}, z) \right] 
   \left( \frac{\Omega_{\mathrm{b}} h^2}{0.023} \right) \nonumber \\
&\times& \left( \frac{0.15}{\Omega_{\mathrm{m}} h^2 \cdot 10} \right)^{1/2} 
   (1+z)^{3/2}\,,
\label{eq:tau}
\end{eqnarray}

where $x_{\rm HI}(\textbf{n},z)$ and $\delta_{\rm b}(\textbf{n}, z)$ are neutral hydrogen fraction and baryonic density contrast respectively along the line of sight direction $\textbf{n}$ and redshift $z$.  We do not consider the effects of the peculiar velocity in this study.

Normally, it is assumed that $\tau({\bf n},z) \ll 1$, and under this assumption the equation (\ref{eq:exact}) can be approximated as, \begin{equation}
\DTB({\bf n},z)=\frac{T_S({\bf n},z) -T_\gamma(z)}{1+z}\tau({\bf n},z),
\label{eq:lin}
\end{equation} 
which can be further written as,
\begin{eqnarray}
    \DTB(\textbf{n},z) = 27 x_{\rm HI}(\textbf{n},z)\big[1+\delta_{\rm b}(\textbf{n}, z)\big]\left( \frac{\Omega_{\rm b}h^2}{0.023}\right)\nonumber \\ \times \left(\frac{0.15}{\Omega_{\rm m}h^2} \frac{1+z}{10}\right)^{1/2} \left[1-\frac{T_{\gamma}(z)}{T_{\rm S}(\textbf{n},z)}\right]   \, \mathrm{mK}.
\label{eq:approx}    
\end{eqnarray} 
The above equation is widely used for modelling the signal, both analytically and numerically. However, the optical depth can be very large at high redshifts in places where the spin temperature is low and the \HI density is large. During CD, the spin temperature can be very low before the onset of X-ray heating when it is coupled to the IGM kinetic temperature. This leads to a large \HI optical depth, particularly in regions with high \HI density. 
For larger values of $\tau$, it is necessary to include higher-order terms, such as $\tau^2$, $\tau^3$, and beyond, in equation (\ref{eq:lin}) to accurately compute $\delta T_{\rm b}(\textbf{n},z)$. 
However, for moderate $\tau$ values equation (\ref{eq:exact}) can be approximated as, 
\begin{equation}
  \DTB(\textbf{n}, z)=\frac{T_S(\textbf {n}, z)-T_{\gamma}(z)}{1+z}\left[\tau-\tau^2/2+\tau^3/6 - \mathcal{O}(\tau^4)\right].
  \label{eq:3rd_ord}
\end{equation} 
This inclusion of higher-order terms would influence various statistics of the signal.

\subsection{Simulations of {\HI} 21-cm signal during CD}
\label{sec:2.2} 
We use {\sc grizzly} code \citep{Ghara_2015,Ghara_2015b,ghara18} for $21$-cm signal simulations, which is based on a one-dimensional radiative transfer scheme. However, recently \citet{Ma_2023, Acharya_2025} developed {\sc polar}, an algorithm that combines the radiative transfer code {\sc grizzly} with the semi-analytical galaxy formation code {\sc L-Galaxies 2020}, providing a self-consistent framework to model galaxy formation, reionization, and the resulting 21-cm signal. {\sc grizzly} requires a gas density field, a halo catalog and a source model as its inputs. The dark matter distribution in a cubic box of comoving length $714.29$ Mpc is obtained using an N-body code named {\sc cubep$^3$m} \citep{Harnois_Deraps_2013} which is smoothed into 600 uniform grids  resulting in a spatial resolution of $1.19$ Mpc. Dark matter distributions are obtained at multiple redshifts in the range of $z=18$ to $z=10$ during CD. Dark matter halos are identified using an on-the-fly halo finder which uses the spherical overdensity method. The lowest dark matter halos identified in our simulations have masses of approximately $10^9$ $\MSUN$ (the mass of a dark matter particle in our simulations is $\sim 10^8$ $\MSUN$).  The stellar mass available within a halo is directly related to the DM halo mass as $M_*=f_* (\frac{\OmegaB}{\Omegam}) M_{\rm h}$, where the star formation efficiency $f_*$ is assumed to be $0.03$ \citep{Behroozi_2015,Sun_2016}. Our simulation has two different astrophysical parameters i.e., $f_{\rm x}$ and $\MHMIN$.   $\MHMIN$ is the minimun mass of a dark matter halo capable of hosting sources that emit $\LYA$, X-ray and UV ionizing photons. We consider two different values of
$\MHMIN$, i.e. $10^9$ $\MSUN$ and $10^{10}$ $\MSUN$, in this study. $f_{\rm x}$ is the X-ray heating efficiency parameter which determines the X-ray emission rate per unit stellar mass, set as $f_{\rm x} \times 10^{42}\ \mathrm{s}^{-1}\ \MSUN^{-1}$. We have considered 3 different values of $f_{\rm x}$, i.e. $0.1,46.4$ and $1000$ in our study.
The details of these simulations are discussed in \citet{Kamran_2022}.

We see in eq. \ref{eq:approx} that we need to estimate $T_S$ in each grid of simulation cube in order to simulate the \HI 21-cm signal i.e., $\delta T_b$.  The spin temperature $T_S$ can be estimated using the equation
\begin{equation}
T_S^{-1}=\frac{T_{\gamma}^{-1}+x_{\alpha}T_k^{-1}}{1+x_{\alpha}}, 
\label{eq:spintemp}
\end{equation}
where $T_k$ is kinetic temperature of the IGM and  $x_{\alpha} $ is the Lyman-$\alpha$ coupling coefficient which can be calculated using the Lyman-$\alpha$ flux $J_{\alpha}$ as,
\begin{equation}
   x_{\alpha}=\frac{16 \pi^6 T_* e^2f_{12} J_0(|r|)}{27 A_{10}T_k m_e c}.
\end{equation}
Here, $J_0$ is the Ly$\alpha$ flux density at distance $r$ from the source. $T_*=h \nu_{21}/k_b=0.068$ K. $J_0(|r|)$ is calculated around each dark matter halo capable of hosting luminous sources. We consider stellar photons bluewards of Ly$\alpha$ which redshifted into the Ly series resonant lines as the dominant source of Ly$\alpha$ photons.  $J_0$ is assumed to scale as $|r|^{-2}$. In overlap regions we add up the contributions of Ly$\alpha$ flux from all sources. The ionization and kinetic temperature profile around each DM halo has been calculated using a set of coupled differential equations (eq. 1 - 3 and 12 of \citet{thomas08}). One of the important ingredients in generating profiles of Ly$\alpha$ flux, hydrogen ionization and kinetic temperature is the spectral energy distribution (SED). Details regarding  modeling of the SED is described in \citet{Ghara_2015}.

 The 21-cm signal simulations start by solving these one-dimensional coupled differential equations and creating many radial profiles of $\XHII$, $\TK$ and $x_{\alpha}$ around isolated sources for different combinations of halo masses, UV emissivity, the ratio of X-ray and UV luminosities, redshifts and density contrast.  In a post-processing step, these profiles are used to generate the ionization maps and kinetic temperature fields at different redshifts. The related steps for producing \HII maps are estimating and assigning the volume of \HII regions around each source, computing the ``unused'' ionizing photons in the overlapped regions, and redistributing the unused photons among the overlapping sources by increasing the size of the \HII regions appropriately. The code then computes $\XHII$ in the partially ionized regions using a simple overlap prescription. Finally, the kinetic temperature maps are generated using a correlation between $\XHII$ and $\TK$ \citep[for details, see][]{Ghara_2015,Ghara_2015b,ghara18, 2019MNRAS.487.2785I}. 
 The code then generates $\TS$ maps using the previously generated $\TK$ maps and $x_{\alpha}$ using eq. \ref{eq:spintemp}, assuming the collisional coupling is negligible at these redshifts.

Using the DM density distribution obtained from the N-body simulation and spin temperature, neutral hydrogen distribution from the one-dimensional radiative transfer code {\sc grizzly}, we also generate maps of the \HI 21-cm optical depth $\tau({\bf n},z)$ distribution. We generate $\delta T_b$ maps at $23$ different redshifts covering the entire CD using both the approximated (equation (\ref{eq:lin})) and the exact (equation (\ref{eq:exact})) $\DTB$ equations. 

\begin{figure}
\includegraphics[width=0.48\textwidth]{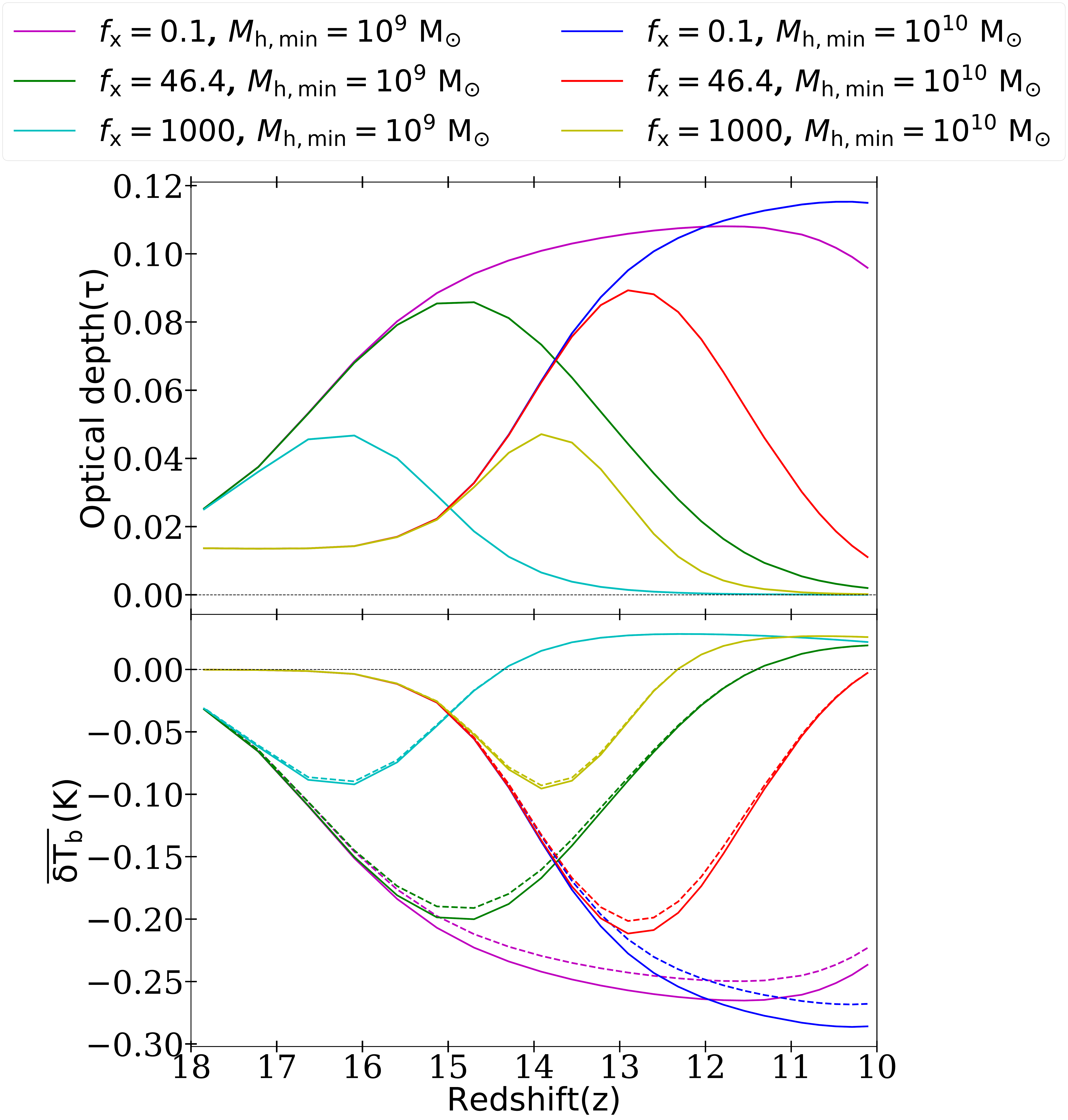}\caption{
Redshift evolution of the mean HI 21-cm optical depth and the differential brightness temperatures. The upper and lower panels show the redshift evolution of mean \HI 21-cm optical depth, $\tau$, and the global 21-cm differential brightness temperature $\overline{\DTB}(z)$, respectively. In the lower panel,  the evolution of both approximated (solid curve) and exact $\overline{\DTB}(z)$ (dashed curve) is shown.
}
    \label{fig: mean}
\end{figure}

We show the evolution of the global signal ($\overline{\DTB}(z)$) with redshift in the lower panel of Fig. \ref{fig: mean}. We see, for all choices of $\FX$ values and for a fixed $\MHMIN$, that the initial trend in the evolution of $\overline{\DTB}(z)$ around redshift $z= 18$ is very similar as only $\LYA$ coupling dictates the 21-cm signal during the initial phase. It helps to couple the spin temperature with the IGM kinetic temperature; therefore, the signal is seen in absorption. Later, X-ray heating becomes important and raises the IGM kinetic and spin temperature.  This creates a trough-like feature in the redshift evolution of $\overline{\DTB}(z)$.   Further, we see that the absorption trough becomes deeper for smaller $\FX$ values as the IGM heating starts late and the spin temperature, which is coupled to IGM kinetic temperature through $\LYA$ coupling, continues to decrease due to adiabatic expansion of universe. Moreover, the trough is more flattened for smaller $\FX$ values. This is because the X-ray heating rate and cooling rate due to cosmological expansion remain similar for a longer period of time in these scenarios. As expected, higher $\FX$ makes $\overline{\DTB}(z)$  positive at higher redshift. The trough like feature in $\overline{\DTB}(z)$ shifts towards lower redshifts for higher halo mass cut-off, mainly because massive DM halos are formed later and, therefore, processes such as $\LYA$ coupling, X-ray heating get delayed.

From the lower panel we also see that if we vary the X-ray heating efficiency $\FX$, both the width   and depth of the troughs vary. These are maximum for $\FX=0.1$ and minimum for $\FX=1000$. Further, we see that $\overline{\DTB}(z)$ calculated using the exact equation is smaller than the approximated method, and the difference is more for smaller values of $\FX$. This is expected as $T_{\rm S}$ has lower values for the $\FX=0.1$ model, i.e., $\tau$ has higher values, which leads to a greater difference in the troughs between approximated and exact $\overline{\DTB}(z)$ compared to the $\FX=1000$ model. The difference is found to be similar for different $\MHMIN$ models for a given $\FX$ value.  The percentage difference between approximated and exact $\overline{\DTB}(z)$ reaches up to $6.2\%$ for [$\FX=0.1$, $\MHMIN=10^9$\ $\MSUN$]  model and $6.7\%$ for [$\FX=0.1$, $\MHMIN=10^{10}$\ $\MSUN$] model. The differences reduce to roughly $5\%$ and $2.7\%$ for $\FX= 46.4$ and $1000$, respectively, for both the $\MHMIN$ values. This difference arises because of considerably large values of $\tau$ during the CD. The higher-order terms in $\exp(- \tau)$ become significant when $\tau$ is large. To elaborate on this, we plot the evolution of the mean $\tau$ with redshift in the upper panel of Fig. \ref{fig: mean}. The redshift at which the mean
$\tau$ reaches its maximum corresponds to the largest difference between the exact and approximated $\overline{\DTB}(z)$. The optical depth becomes maximum when the IGM kinetic  and spin temperature are minimum, which occurs just before the onset of X-ray heating.

\section{Results}
\label{sec:3}
This section presents results on the distributions of HI 21-cm optical depth and differential brightness temperature. It  further shows results on the redshift evolution of skewness and bispectrum, as well as the impact of large optical depth on these quantities.

\subsection{{\HI} 21-cm optical depth ($\tau$) distribution}
\label{sec:3.1}

\begin{figure}
\includegraphics[width=0.24\textwidth]{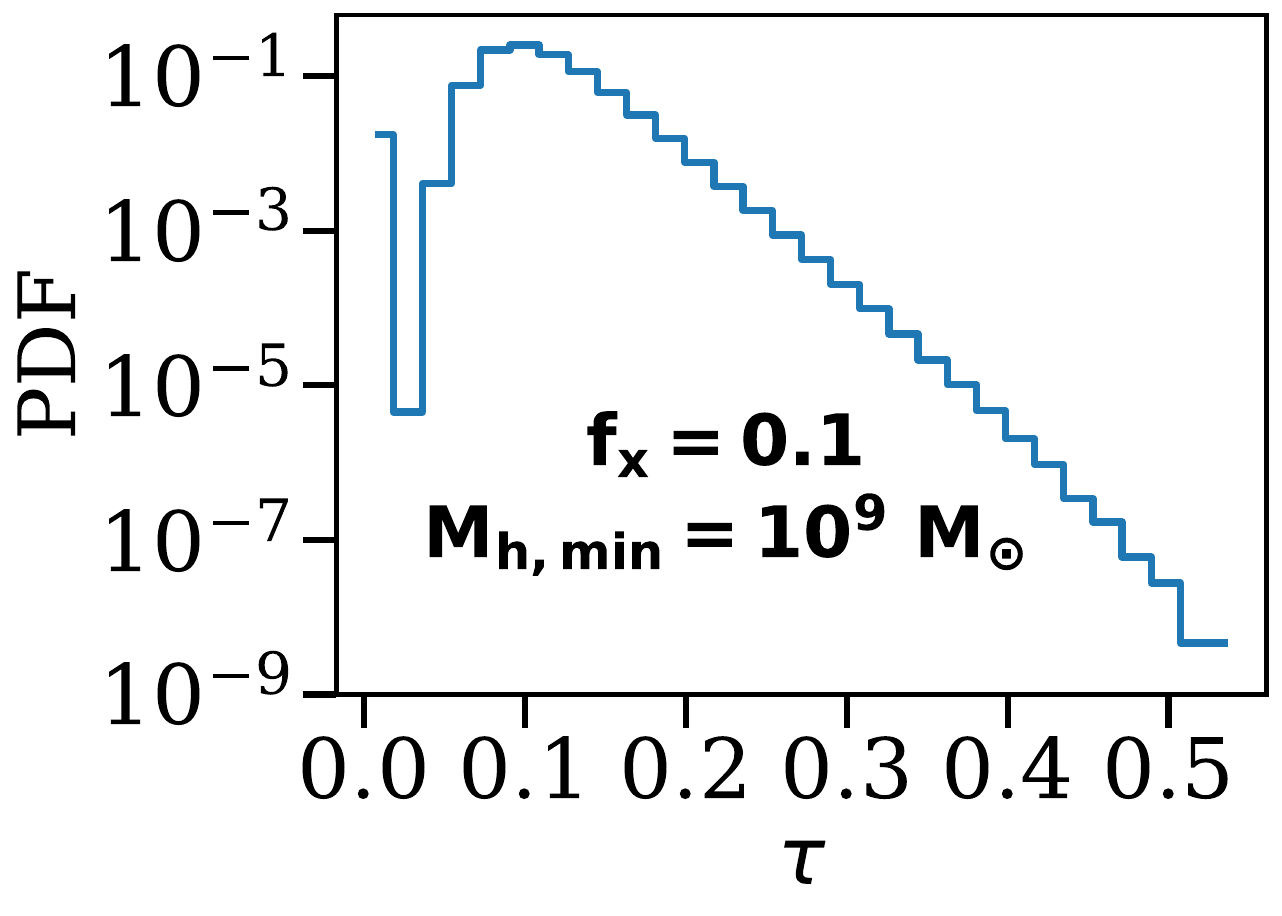}
    \includegraphics[width=0.25\textwidth]{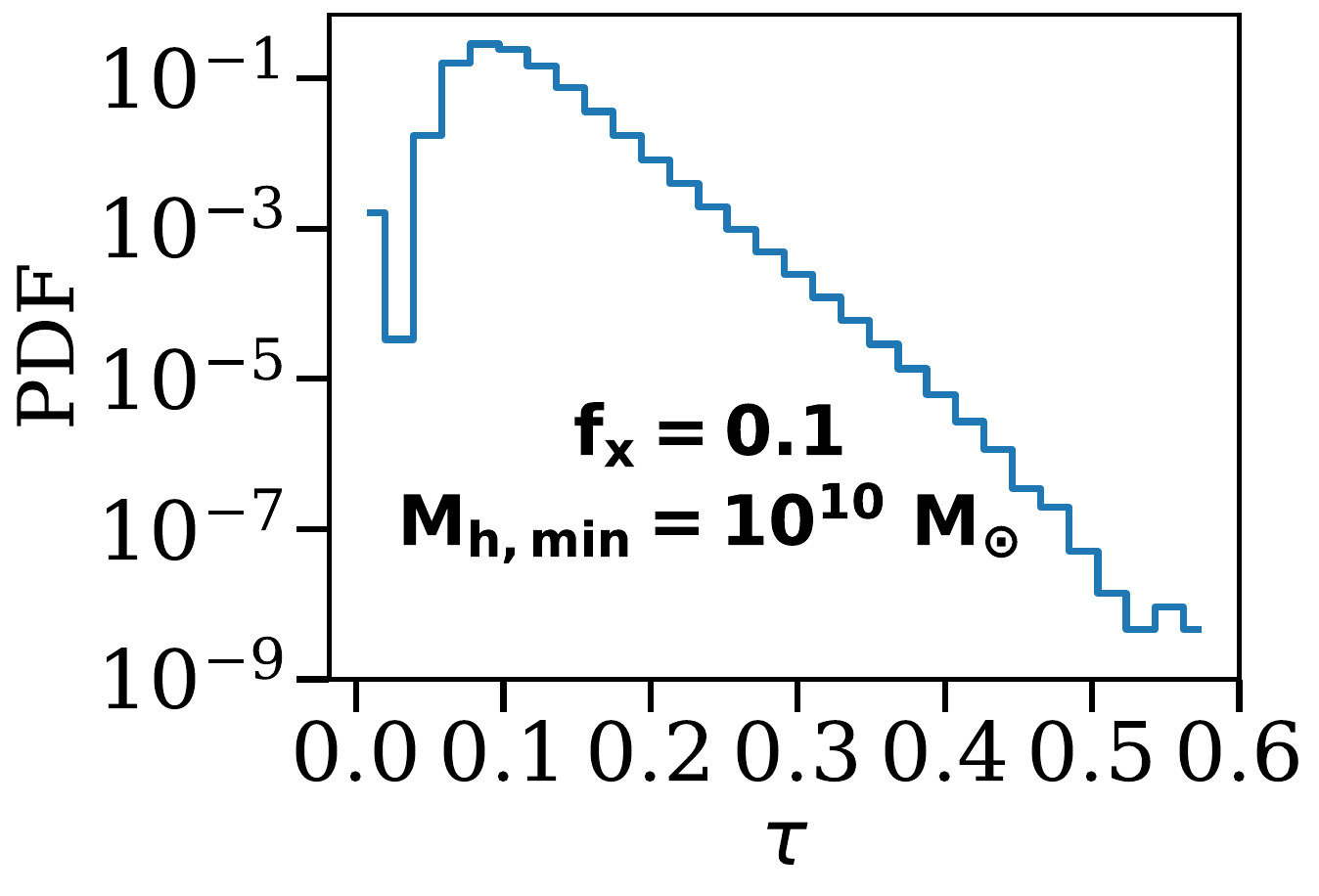}
    \includegraphics[width=0.245\textwidth]{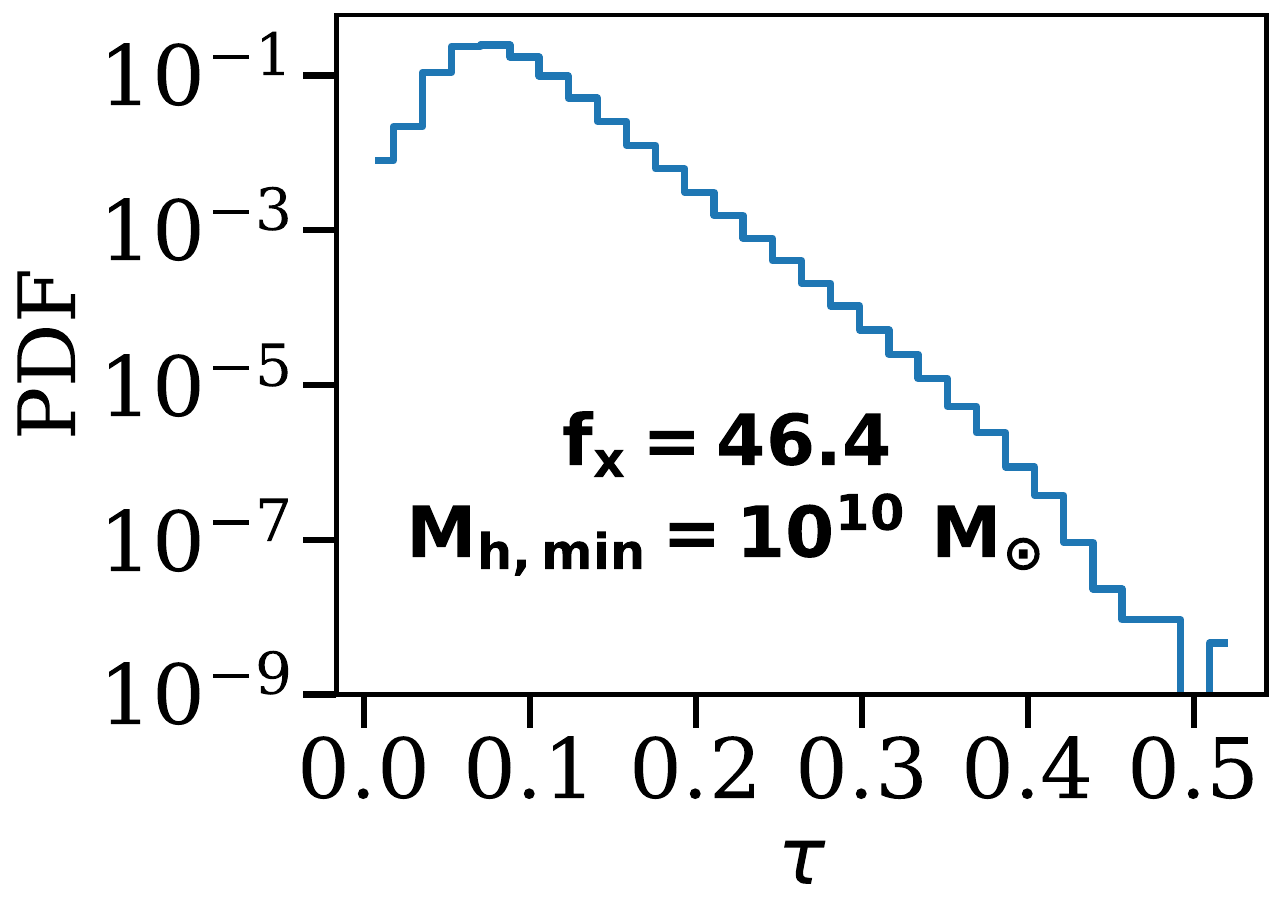}
   \includegraphics[width=0.235\textwidth]{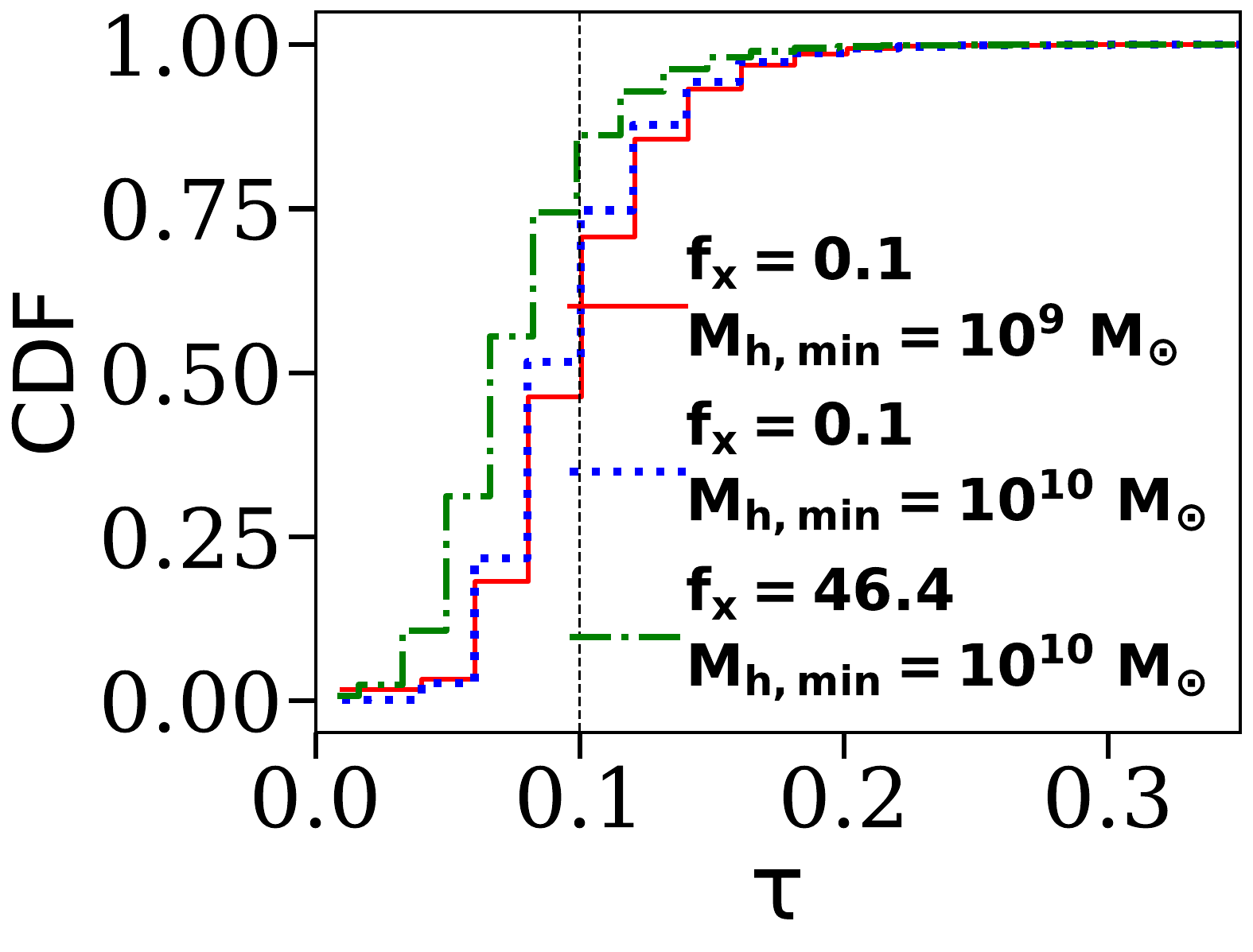}
    
    \caption{Probability distribution function (PDF) of \HI 21-cm optical depth ($\tau$) for models with $\FX=0.1$, $\MHMIN=10^9$\ $\MSUN$ (top left);  $\FX=0.1$, $\MHMIN=10^{10}$\ $\MSUN$ (top right) and $\FX=46.4$, $\MHMIN=10^{10}$\ $\MSUN$ (bottom left), at redshift $z=12.3$. The bottom right panel shows the cumulative distribution function (CDF) of $\tau$ for all three scenarios.}
    \label{fig:tau-hist}
\end{figure}

We plot the probability distribution function (PDF) and the cumulative distribution function (CDF) of simulated $\tau$ at redshift $z=12.3$ for three  models in Fig. \ref{fig:tau-hist}. The average $\tau$ values at $z=12.3$ for these models are  higher compared to other redshifts, which can be seen in the upper panel of Fig. \ref{fig: mean}. 
We see from the CDF (refer to the bottom right panel of Fig.\ref{fig:tau-hist}) that as many as $\sim 55\%$ of pixels in the simulated cube have $\tau$ value greater than $0.1$ for the model corresponding to $\FX = 0.1$, $\MHMIN = 10^9\ \MSUN$. For models with $\FX = 0.1$, $\MHMIN = 10^{10},\MSUN$ and $\FX = 46.4$, $\MHMIN = 10^{10},\MSUN$, about $49\%$ and $25\%$ of the total pixels, respectively, have $\tau > 0.1$. Furthermore, we observe that in all scenarios, the distribution of $\tau$ is positively skewed. The optical depth ($\tau$) depends on both the matter density and the spin temperature (refer to equation (\ref{eq:tau})), both of which are non-Gaussian. This results in a non-Gaussian $\tau$ field, and consequently contributes to a non-Gaussian $\delta T_b$ field during CD.

\subsection{Non-Gaussian \HI \textbf{21-cm signal}}
\label{sec:3.2}
\begin{figure}
\includegraphics[width=0.485\linewidth]{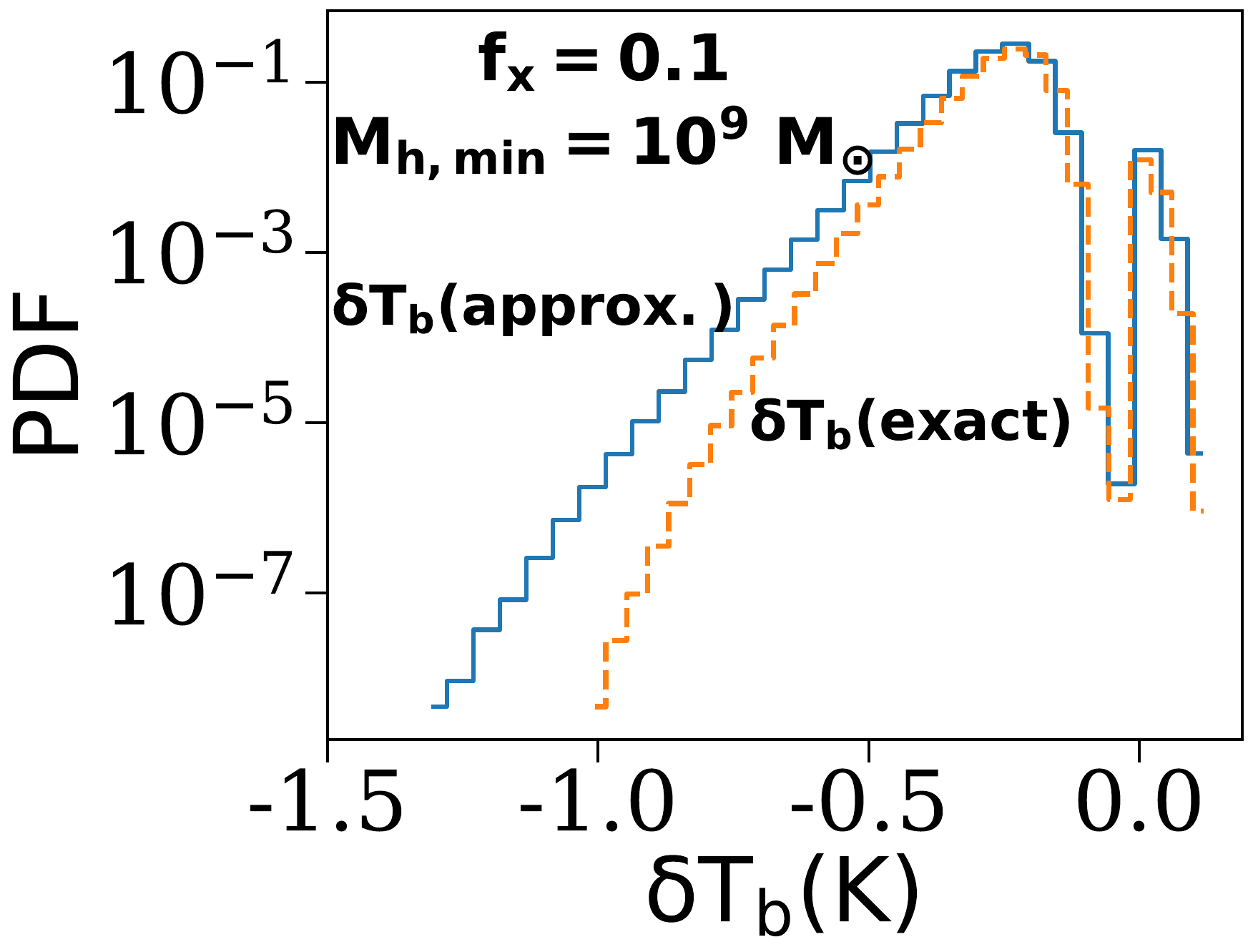}
\includegraphics[width=0.485\linewidth]{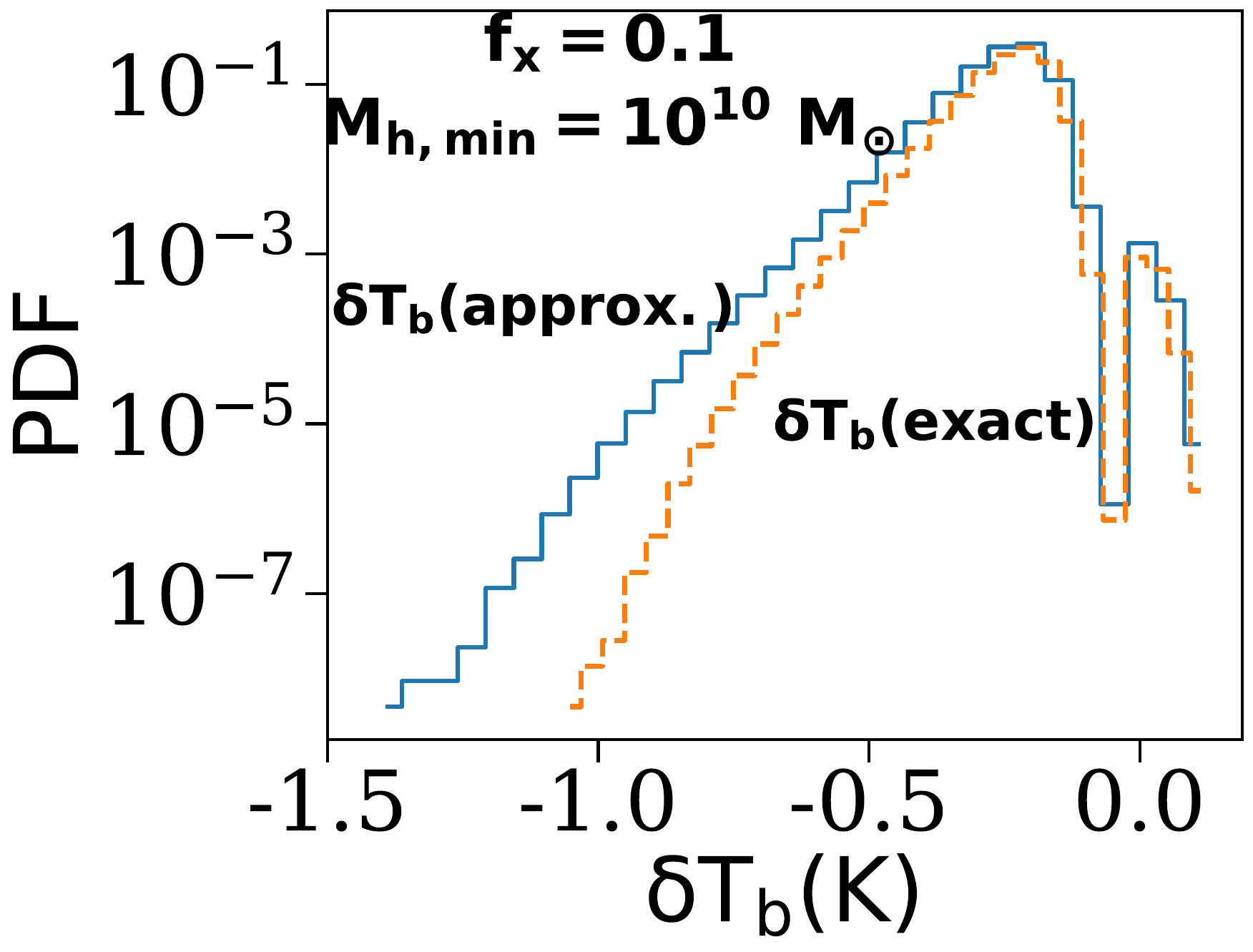}
\centering
\includegraphics[width=0.49\linewidth]{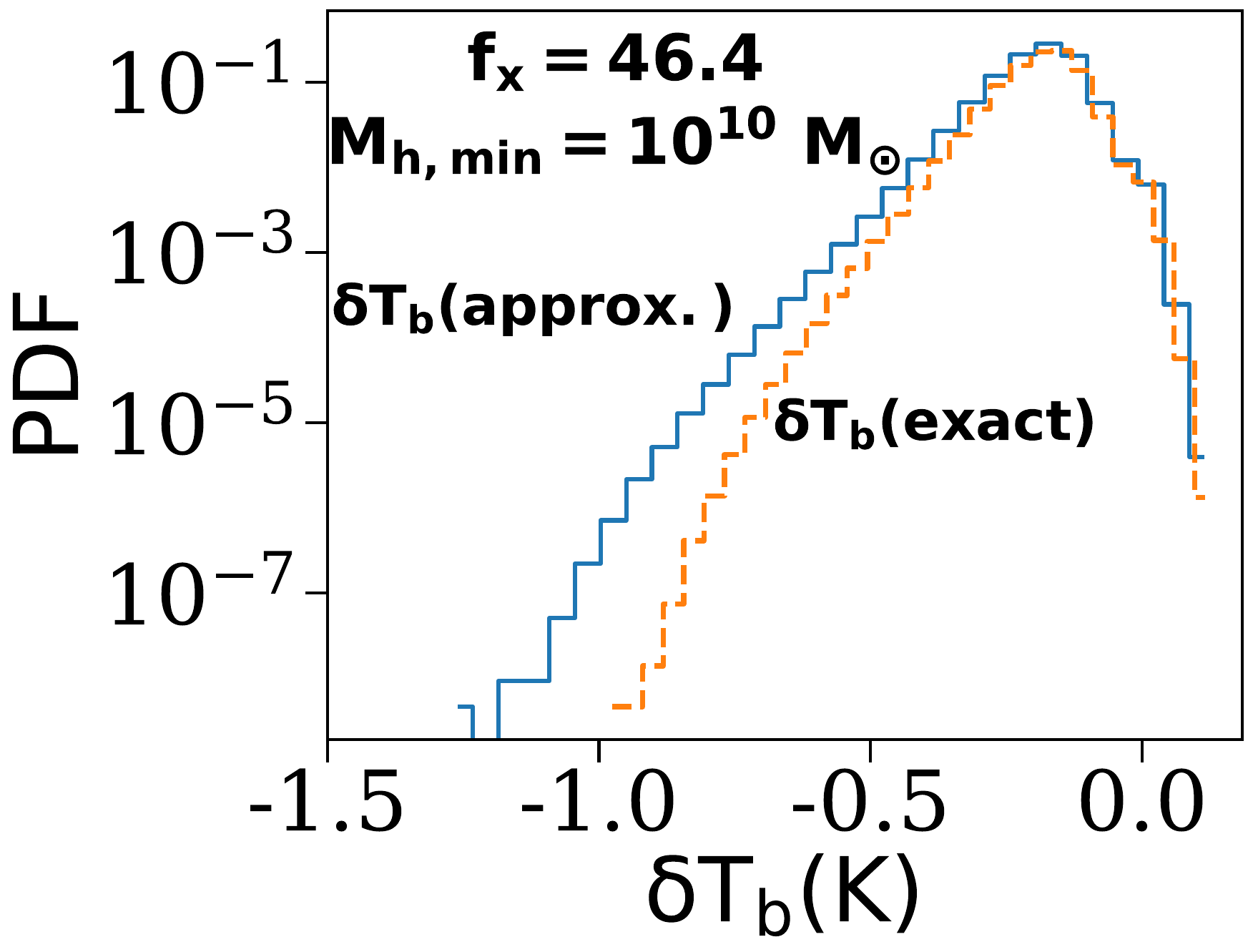}
    \caption{Probability distribution function (PDF) of simulated $\DTB$ for models with $\FX=0.1$, $\MHMIN=10^9$\ $\MSUN$;  $\FX=0.1$, $\MHMIN=10^{10}$\ $\MSUN$ and $\FX=46.4$, $\MHMIN=10^{10}$\ $\MSUN$, at redshift $z=12.3$. 
    }
    \label{fig:PDF}
\end{figure}

Fig.\ref{fig:PDF} shows the probability distribution functions (PDF) of $\DTB$ estimated using the approximated and exact equations for three different scenarios of CD. For these models we find that $\DTB$ distribution is non-Gaussian and negatively skewed (Fig.\ref{fig:skew}) at $z\approx12.3$. We further see that the exact $\DTB$ is less skewed compared to the approximated one. The reason is that, as revealed in our simulations, $\tau$ can become considerably large during the Cosmic Dawn. Therefore, higher-order terms like $\tau^2$ and $\tau^3$ in $\delta T_b$ (refer to eq. \ref{eq:3rd_ord}) can no longer be neglected. An accurate calculation of $\DTB$ reduces its skewness compared to the approximated calculation due to the presence of higher-order terms of $\tau$ in the exact equation. We further notice that $\DTB$ with large negative values are more suppressed, as these $\DTB$s are associated with large $\tau$ values, and therefore, are more affected.  Voxels with small $|\DTB|$ values are less affected as these are associated with very small $\tau$. Therefore,  large optical depth not only suppresses the overall fluctuations in $\delta T_b$ but also impacts the non-Gaussian nature of the signal. In other words, we can say that higher-order terms such as $\tau^2$, $\tau^3$ introduce additional non-Gaussianity to $\DTB$ distribution.

\subsection{Skewness and its redshift evolution}
\label{sec:3.3}
\begin{figure*}
\includegraphics[width=0.95\textwidth]{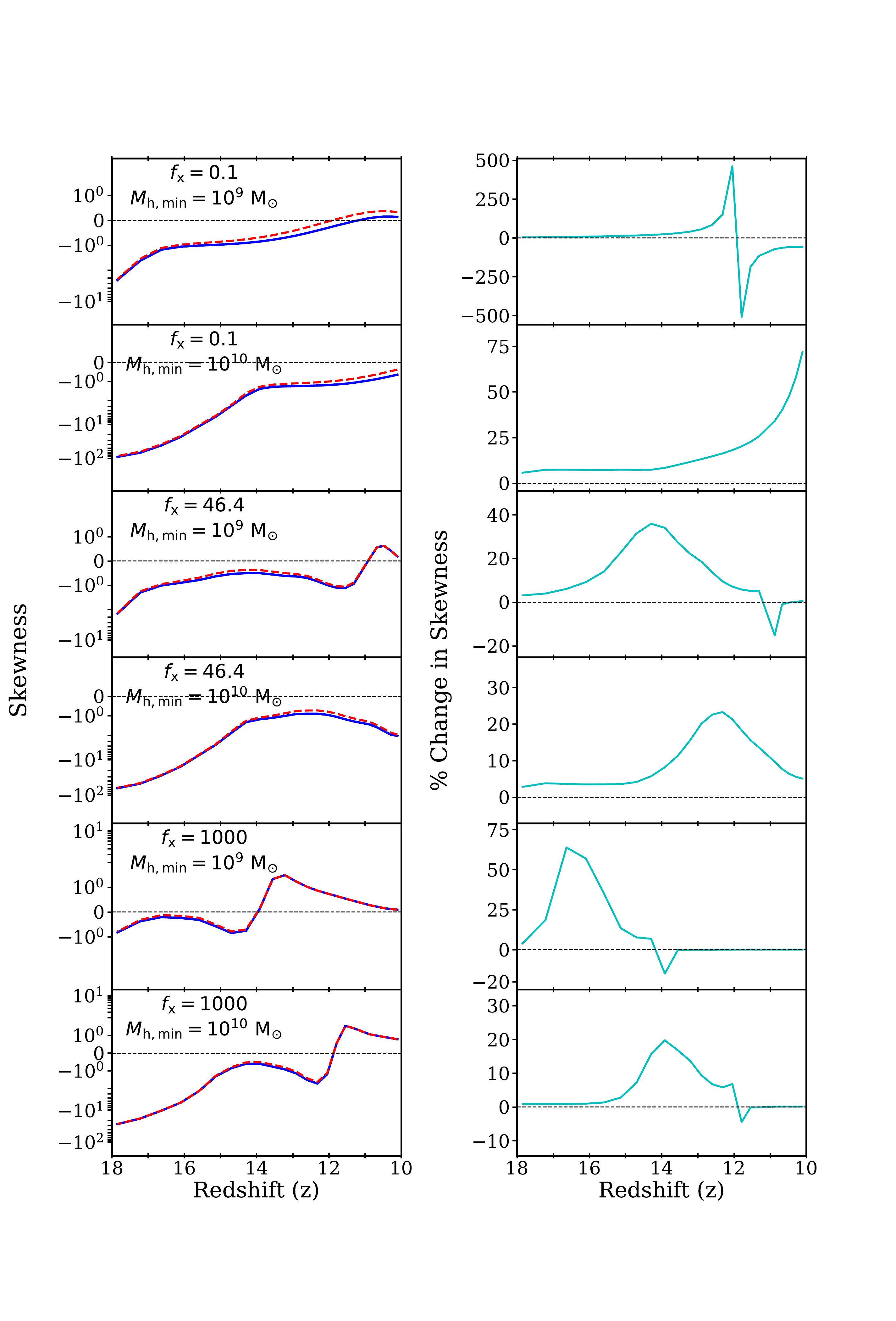}
\vspace{-2.2cm}
    \caption{Redshift evolution of skewness for the approximated and exact $\DTB$ (left panels) and the percentage change (right panels) for all the CD scenarios. In the left panel the blue solid and red dashed lines represent the skewness for the approximated and exact $\DTB$ respectively.}
    \label{fig:skew}
\end{figure*}

The presence of a large numbers of $\LYA$-coupled regions during the early CD,  heated regions during the heating phase, and ionized bubbles during reionization enhances the non-Gaussian signal and its skewness. As $\LYA$ coupling and IGM heating progress toward completion, non-Gaussianity gradually decreases. $\DTB$ being positive inside heated regions makes the distribution positively skewed. On the other hand, it is negative inside the $\LYA$ coupled regions before heating, resulting in a negatively skewed signal.  

In order to quantify non-Gaussianity in the signal and impact of higher-order terms of $\tau$, we estimate the skewness of $\DTB$ as a function of redshift for different scenarios. Skewness is a measure of asymmetry in the PDF from a Gaussian distribution, and the simplest one-point statistic that is sensitive to non-Gaussianity. Skewness for $\DTB$ is defined as,
\begin{equation}
    \rm{skewness} = \left\langle \left\{ \frac{\DTB({\bf n}, z) -{\overline {\rm \delta T}}_b(z)}{\sigma_{\DTB}} \right\}^3 \right\rangle, 
\end{equation}
where $\sigma_{\DTB}$ is the standard deviation of the $\DTB$ field. 

The left column of Fig. \ref{fig:skew} shows skewness as a function of redshift $z$ for all scenarios considered here. We can see that at the beginning of CD, skewness is largely negative around redshift $z = 18$. As redshift decreases, its magnitude gradually reduces until the heating of the IGM begins. However, during the initial stages of X-ray heating the skewness starts becoming more negative again. As redshift decreases further, heated regions around sources grow significantly and percolates through the IGM. As a result, skewness becomes positive and continues to increase. Eventually, when the entire IGM is nearly heated, the skewness decreases again.
 For example, in the model with $\MHMIN=10^{10}$ $\MSUN$ and $\FX=1000$,  skewness is $\approx -30$ at redshift $z \approx 18$.  Its magnitude gradually decreases until redshift $z \sim 14$ as the $\LYA$ coupling spreads through the IGM and nears completion. Following this, X-ray heating starts, and skewness tends towards more negative values until redshift $z \approx 12.5$. However, as the heated regions expand and percolates through the IGM, the skewness quickly becomes positive, reaching a peak around $z \approx 11.5$. Later, skewness decreases again as almost the entire IGM gets heated. A similar pattern is observed in other scenarios. However, the actual timings of these processes differ depending on the values of $\MHMIN$ and $\FX$. In general, higher values of $\MHMIN$ and/or lower values of $\FX$ delay these processes.

We can further see from the left column of Fig. \ref{fig:skew} that the initial trend in the redshift evolution of skewness is almost similar for a fixed $\MHMIN$ value when $\LYA$ coupling dominates. This redshift range is from the beginning up to $z\approx 16.5$ for $\MHMIN=10^9$\ $\MSUN$ and upto $z\sim 14$ for $\MHMIN=10^{10}$\ $\MSUN$ models. For larger value of $\FX$, X-ray heating starts earlier and dominates the $\LYA$ coupling, shortening its redshift window (lower panel of Fig. \ref{fig: mean}).

\subsection{Impact of large optical depth on  Skewness}
\label{sec:3.4}
We now shift our focus to the impact of high \HI 21-cm optical depth on $\DTB$ and its statistics. If we choose a model where heating is inefficient as in $\FX=0.1$, $\MHMIN=10^9$\ $\MSUN$, $\tau$ would be sufficiently large throughout the entire CD as the spin temperature remains very low. As a result, $\DTB$ calculated using exact equation is considerably suppressed compared to $\DTB$ calculated using the approximated equation (see lower panel of fig.\ref{fig: mean}).  Further, the skewness of $\DTB$  is suppressed compared to that estimated using the approximated $\DTB$.

During the initial stages of CD, the skewness of the approximated and exact $\DTB$ takes large negative values. This is because $\LYA$ radiation couples the \HI 21-cm spin temperature to the IGM kinetic temperature within regions surrounded by the sources, whereas the background IGM remains coupled with the CMBR temperature. This makes the \HI 21-cm signal negative inside the sources and consequently the overall signal negatively skewed. We do not see any significant difference between the exact and approximated $\DTB$, mainly because \HI 21-cm optical depth is small during the initial stages of CD. 

However, in Fig. \ref{fig:skew} we see that the magnitude of skewness decreases as we go from redshift $z=18$ to lower redshifts. During this period, the $\LYA$ coupling percolates through the IGM and eventually results into a uniform $\LYA$ coupled  IGM. The average spin temperature decreases and $\tau$ increases until the redshift around which the heating starts. This makes the higher-order terms of $\tau$ in equation (\ref{eq:exact}) important. Consequently, the difference in skewness between the approximated and the exact $\DTB$ also increases with decreasing redshift. This change can go up to $\sim 100\%$ in some cases considered here. 

 Once X-ray heating starts, kinetic and spin temperatures start increasing with decreasing redshift. Consequently, the optical depth starts to decrease with redshift. This makes the contribution of higher-order terms less important. Therefore, the exact $\DTB$ is very similar to the approximated one toward the end of CD.

 Due to very inefficient heating in models corresponding to $\FX=0.1$, the IGM, including the $\LYA$ coupled regions near the sources, remains cold for an extended period of time. Therefore, the average $\DTB$ remains negative and the PDF of $\DTB$ remains negatively skewed up to redshift $z \sim 11$. Cold IGM in these models also lead to a larger optical depth making higher-order terms more impactful on the global signal as well as on the skewness. We see that (first two top-right panels in Fig. \ref{fig:skew}) the differences in skewness go up to $\sim 500 \%$ and $\sim 75 \%$ for models with $(\FX, \MHMIN) =(0.1, 10^9 \, \MSUN) $ and $(0.1, 10^{10} \, \MSUN)$, respectively. 
  
X-ray heating is moderate in the model with $\FX=46.4$, $\MHMIN=10^9 \, \MSUN$. Therefore, the change in the skewness between approximated and exact $\DTB$ goes up to  $\sim40\%$ at redshift $z \sim 14$ when the average optical depth becomes maximum in the model. Similarly, we see that the change is only up to $\approx 25 \%$ at redshift $z \approx 12$ for $\MHMIN=10^{10} \, \MSUN$ for the same X-ray efficiency parameter. 
Models with $\FX = 1000$ exhibit similar features, but these appear at relatively earlier redshifts. Here, the maximum difference in skewness between the exact and approximated models is $\approx 60 \%$ and $\approx 20 \%$ 
 respectively for $(\FX, \MHMIN) =(1000, 10^9 \, \MSUN) $ and $(1000, 10^{10} \, \MSUN)$.

\subsection{Redshift evolution of bispectrum}
\label{sec:3.5}
Bispectrum is the lowest order many-point statistics that is sensitive to non-Gaussianity. Bispectrum of a $\DTB$ field within a volume $V$ can be written as,
\begin{equation}
\delta^{\rm K}_{3\rm D}(\mathbfit{k}_1+\mathbfit{k}_2+\mathbfit{k}_3)~  B(\mathbfit{k}_1,\mathbfit{k}_2,\mathbfit{k}_3)=V^{-1} \langle \Delta_{\rm b}(\mathbfit{k}_1) \Delta_{\rm b}(\mathbfit{k}_2) \Delta_{\rm b}(\mathbfit{k}_3) \rangle\nonumber, 
\end{equation}
where $\Delta_{\rm b}(\mathbfit{k})$ is the Fourier conjugate of $\DTB(\mathbfit{x})$, $\langle \cdots \rangle$ denotes an ensemble average. The Kronecker delta function $\delta^{\rm K}_{3\rm D}$ in the definition is an implication of statistically homogeneity of the signal which ensures that $\mathbfit{k}$-triplets must form a closed triangle for a non zero bispectrum. We computed bispectrum of the simulated $\DTB$ maps using an FFT-based fast estimator \citep{Shaw_2021} where the Kronecker delta function is replaced by summation of plane waves over grid points in real space. To determine the shape of the triangle formed by the three $\mathbfit{k}$-modes, this algorithm uses the parametrization: $\mu=-(\mathbfit{k}_1\cdot \mathbfit{k}_2)/(k_1~k_2)$ and $t=k_2/k_1$, where $k_i=|\mathbfit{k}_i|$. The cosine of the angle between $\mathbfit{k}_1$ and $\mathbfit{k}_2$ is denoted as $\mu$, whereas $t$ is the ratio between them and each set of  $(k_1,\mu,t)$ represents a different triangle configuration. For unique triangle configuration, order of three $k$ sides and the range of $\mu$ and $t$ values are prescribed in \citep{Bharadwaj_2020} as,
\begin{eqnarray}
k_1\geq k_2 \geq k_3;\quad{\rm}
0.5 \le t,\, \mu \le 1 \quad {\rm and}\quad 2 \mu t \ge 1. \ 
\nonumber
\end{eqnarray}
We have divided the $k$-space into $20$ logarithmic spherical shells for bispectrum estimation. Each combination of three shells forms a particular bin in the triangle shape and size space $(k_1, \mu, t)$. The average size of the bin corresponds to $k_1$. Sampling of ($\mu,t$) space for a fixed $k_1$ represents different triangle shape configurations. We suggest the reader to \citet{Bharadwaj_2020} and \citet{Shaw_2021} for the details about the triangle shape and the bispectrum framework. In this work, we consider only the equilateral ($k_1=k_2=k_3$) and the squeezed limit triangles ($\mathbfit{k}_1 \approx -\mathbfit{k}_2$, $\mathbfit{k}_3 \approx 0$) for our bispectrum analysis.

\begin{figure*}
\centering
\vspace{-2.0cm}
\includegraphics[width=0.85\textwidth]{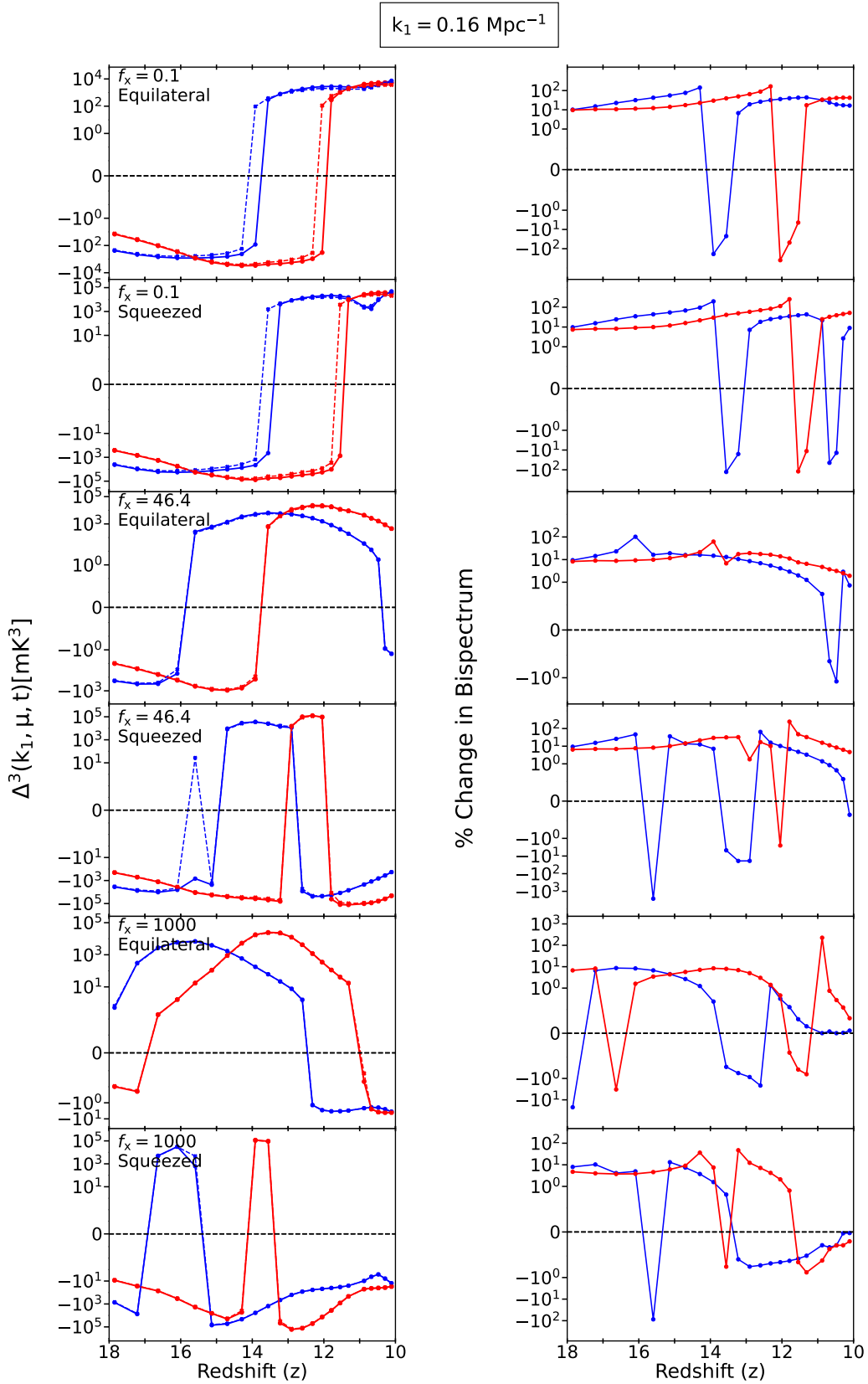}
\vspace{-2.8cm}
\caption{Redshift evolution of the spherically averaged dimensionless bispectrum for approximated and exact $\DTB$, along with the percentage change between them,  at the scale $k_1=0.16\, {\rm Mpc}^{-1}$. In the left panel, the blue solid and dashed lines represent bispectrum of approximated and exact $\DTB$, respectively, for $\MHMIN = 10^9 \MSUN$ while, the red solid and dashed lines represent the same for $\MHMIN = 10^{10} \MSUN$. In the right panel, the blue and red solid lines represent the percentage difference between the approximated and exact $\DTB$ for the models with $\MHMIN = 10^9\ \MSUN$ and $\MHMIN = 10^{10}\ \MSUN$, respectively.}
\label{fig:bs1}
\end{figure*}

\begin{figure*}
\includegraphics[width=0.86\textwidth]{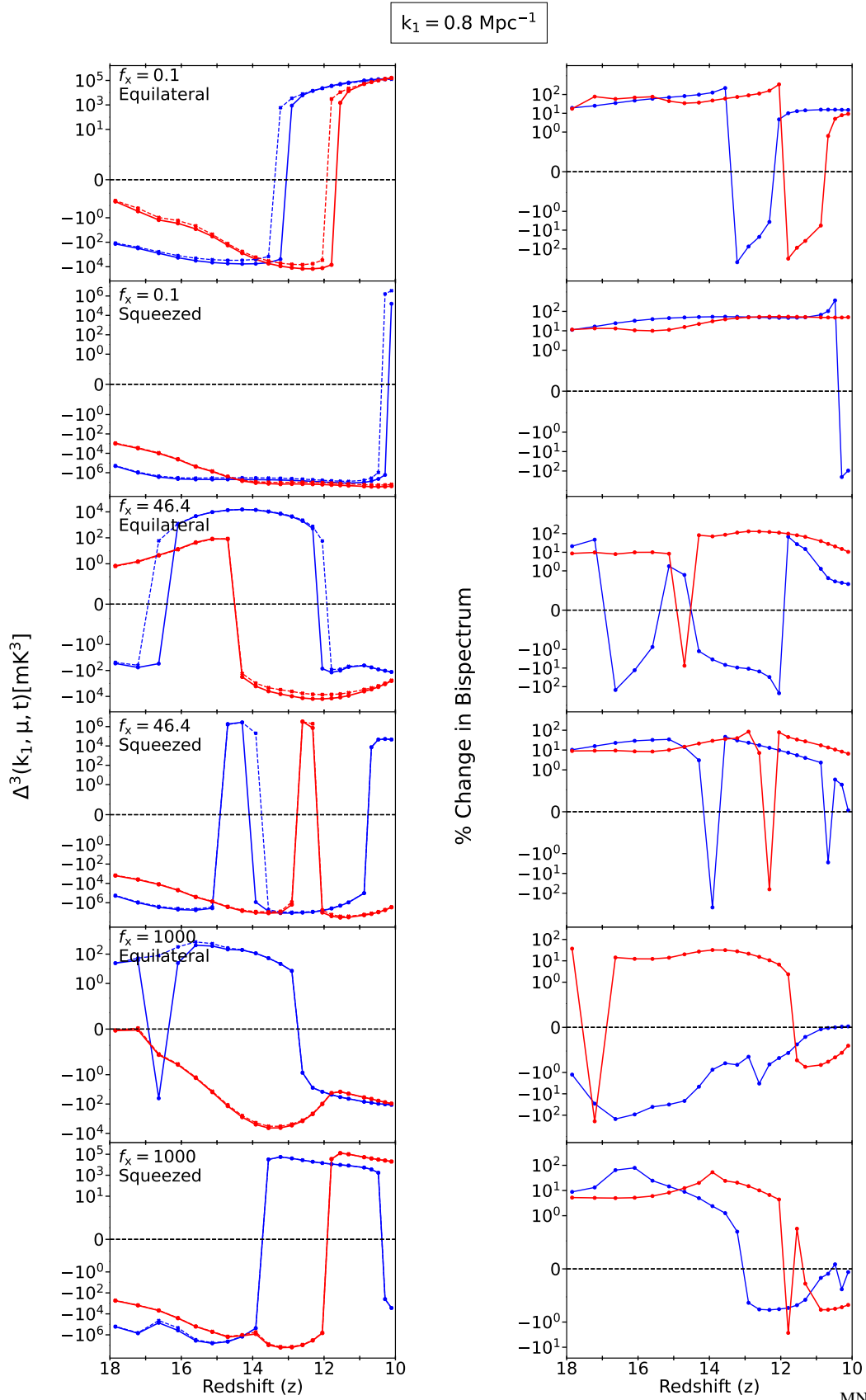}
\vspace{-3cm}
\caption{ Same as Fig. \ref{fig:bs1} for the scale $k_1=0.8\, {\rm Mpc}^{-1}$.}
\label{fig:bs2}
\end{figure*}

In Figs. \ref{fig:bs1} and  \ref{fig:bs2}, we show the redshift evolution of bispectra for both the exact and approximated $\DTB$ at two different length scales. Fig. \ref{fig:bs1} shows results for a large length scale (small $k$) corresponding to $k_1=0.16$ Mpc$^{-1}$ and Fig. \ref{fig:bs2} shows results for a small length scale corresponding to $k_1=0.8$ Mpc$^{-1}$. We plot the spherically averaged dimensionless bispectrum $\Delta^{3}(k_1,\mu,t)={k_1}^3{k_2}^3B(k_1,k_2,k_3)/(2\pi^2)^2$ \citep{Majumdar_2020} for both the equilateral and squeezed limit triangles. The bispectrum corresponding to the equilateral triangles finds the correlation between three equal length scales. In contrast, the bispectrum corresponds to the squeezed limit configuration calculates the correlation between two equal length scales and one very large length scale.

In  Fig. \ref{fig:bs1},  we see that $\Delta^{3}(k_1,\mu,t)$ is initially negative for $\FX = 0.1$ and $46.4$. At the beginning of CD, the spin temperature $T_{S}$ remains coupled to the CMBR temperature $T_\gamma$. This makes the \HI 21-cm signal nearly undetectable. However, as the first sources emerge, $\LYA$  photons start emitting from these sources as a result, $\LYA$ coupled regions from these sources appear as regions of negative $\DTB$  against a zero background signal. This results in a negative bispectrum that continues as long as the $\LYA$ coupled regions remain isolated. However, later these regions expand and percolate through the IGM. This widespread $\LYA$ coupling ultimately reduces the non-Gaussianity of the \HI 21-cm signal. For $\FX=0.1$, we find that the bispectrum is negative (Fig. \ref{fig:bs1}) from the start of CD until the redshifts $z \approx 14$ and $\approx 12$ for $ \MHMIN=10^9 $ and $10^{10} \, \MSUN$, respectively, both for the equilateral and squeezed triangle configurations. For a fixed $\FX$ value, the overall nature in the features of bispectrum for both halo mass scenarios is almost same. The only difference is that, for higher $\MHMIN$ value, all the processes start at lower redshift.  For higher values of $f_x$, bispectrum turns positive earlier, as can be seen in Fig. \ref{fig:bs1}. However, we note that the bispectrum is positive since the beginning of CD for $\FX=1000$ and the equilateral triangle configuration. 

As the $\LYA$ coupling approaches completion, X-ray radiation from the first sources begins to heat the surrounding IGM. The presence of large heated regions around the sources, buried in a colder IGM, makes the bispectrum positive. We see, in Fig. \ref{fig:bs1}, that the bispectrum becomes positive at redshifts $z \approx 14$ and $\approx 12$ for $\FX = 0.1$, corresponding to $\MHMIN = 10^9 \MSUN$ and $10^{10} \MSUN$, respectively. The transition redshifts from negative to positive bispectrum are $z \approx  16$ and $\approx 14$  for $\FX = 46.4$ and $\MHMIN = 10^9 \MSUN$ and $10^{10} \MSUN$, respectively, in the equilateral triangle configuration. The transition occurs slightly later, at $z \approx 15$ and $\approx 13$, for the squeezed triangle configuration with the same halo mass thresholds. For $\FX = 1000$, X-ray heating is highly efficient, and the bispectrum remains positive from the beginning of CD in the equilateral triangle configuration. However for the squeezed triangle configuration, the bispectrum is initially negative for a short period of time, and becomes positive as X-ray heating starts. We see in this study that, for squeezed limit configuration, all the features appear at later stages compared to the equilateral one. The reason is that, in the squeezed-limit triangle configuration, there is a correlation involving one very large length scale, and astrophysical processes begin to affect such large scales at later times, i.e., at lower redshifts.

As the X-ray heating progresses, it gradually percolates through the IGM. The widespread heating again reduces the bispectrum. We see this effect in Fig. \ref{fig:bs1} for $\FX = 46.4$, for both equilateral and squeezed triangle configurations. Toward the final stage of heating, almost the entire IGM becomes heated, except  a few underdense regions that are located far from the sources. This creates a scenario where $\DTB$ is positive in most regions of the IGM but remains negative in those cold and underdense isolated pockets. This results in a large-scale negative bispectrum. For example, for $\FX=46.4$ and $1000$, we see that the bispectrum becomes negative again at later times.

For $\FX = 1000$, X-ray heating is highly efficient. We find that the bispectrum becomes positive from the beginning of CD in the equilateral triangle configuration. Later, it goes negative at $z \approx 12.5$ and $\approx 10.5$ for $\MHMIN = 10^9 \MSUN$ and $10^{10} \MSUN$, respectively. However, for the squeezed triangle configuration, the bispectrum is initially negative and then becomes positive as the X-ray heating starts. Finally, it returns to negative value at relatively higher redshifts at $z \approx 15$ and $\approx 13$ for $\MHMIN = 10^9 \MSUN$ and $10^{10} \MSUN$, respectively, for the same reason as stated above.

Figure \ref{fig:bs2} shows the bispectra for $k_1 = 0.8 \, {\rm Mpc}^{-1}$. The overall trends for $\FX=0.1$ and $46.4$ are similar to those seen for $k_1 = 0.16 \, {\rm Mpc}^{-1}$ (Fig. \ref{fig:bs1}). However, there are some notable differences. For the squeezed triangle configuration and $\FX=46.4$, the transition from a negative to a positive bispectrum happens much later—around redshift  $z \sim 10.5$. 
Another distinct feature at  $k_1=0.8 \, {\rm Mpc}^{-1}$ is seen for $\MHMIN=10^{10} \, \MSUN$, $\FX=1000$, and the equilateral triangle configuration. Here, the bispectrum remains negative throughout the entire CD period. Finally, for the squeezed triangle configuration with $\FX=1000$, the transition from negative to positive bispectrum also occurs much later compared to the case of $k_1=0.16 \, {\rm Mpc}^{-1}$.

\subsection{Impact of large optical depth on bispectrum}
\label{sec:3.6}
The right columns of Figures \ref{fig:bs1} and \ref{fig:bs2} show the corresponding percentage change in the exact bispectra compared to the approximated bispectra calculated as, $\frac{\Delta^3_{\rm approx}-\Delta^3_{\rm exact}}{\Delta^3_{\rm exact}}\times100\%$ as a function of redshift. We observe that, for most of the CD period, the magnitude of the exact bispectrum is suppressed compared to the approximate one. This suppression is mainly caused by the term $- \tau^2/2$ in the expression for the differential brightness temperature (equation \ref{eq:3rd_ord}), which reduces the value of $\DTB$.

Since $\tau^2 \propto \rho^2_{\rm HI}/T^2_S$, the suppression is not uniform across the
field. It is  stronger in regions with a higher neutral hydrogen density and lower spin temperature. As a result, the $\tau^2/2$ term affects the non-Gaussian signal (i.e., the bispectrum and skewness) more strongly than it does the global 21-cm signal or the power spectrum.  For example, the percentage change in the bispectrum is found to be $\sim 10\%$ at redshift $z=18$ across all models and triangle configurations. As redshift decreases, this change increases, reaches a maximum at an intermediate redshift, and then declines again. This trend can be explained if we look at the redshift evolution of \HI 21-cm optical depth.  As shown in Fig.\ref{fig: mean}, the average optical depth increases, peaks at a certain redshift, and then decreases at lower redshifts. The percentage difference in the bispectrum becomes largest when the optical depth is highest, consistent with what is observed in the right columns of Figs. \ref{fig:bs1} and \ref{fig:bs2}.  We find that the percentage change in the bispectrum increases up to $\sim 250 \%$ at redshifts $z \approx 14$ and $\approx 12$ for $\FX=0.1$, for both the equilateral and squeezed triangles at $k_1 = 0.16 \, {\rm Mpc}^{-1}$. A similar trend is observed on smaller scales, e.g., at $k_1 = 0.8 \, {\rm Mpc}^{-1}$ (see Fig. \ref{fig:bs2}) where the change reaches up to $\sim 300\%$. This large change is due to the fact that both the IGM kinetic temperature and hence the spin temperature remain very low in this case because heating is highly inefficient. In models with moderate heating ($\FX=46.4$), the percentage change is smaller, typically lies between $\sim 10\%$ to $\sim 200\%$. This is because heating is more efficient than in the low $\FX$ case. This results in higher IGM and spin temperatures. As a result, the \HI optical depth becomes lower which reduces the impact of higher-order terms in $\tau$. For models with very efficient heating ($\FX=1000$), the IGM and spin temperatures are large from the beginning of the CD. This makes the contribution of higher-order terms less significant. Consequently, the change in the bispectrum is small, around $10 \%$, in these cases. However, we find that the change reaches up to $\sim 200\%$ at a few redshifts. We also find that the percentage change can sometimes be negative. This typically occurs when the bispectrum transitions from negative to positive or vice versa.

\section{Impact on power spectrum}
\label{sec:4}
\begin{figure*}
\includegraphics[width=\textwidth]{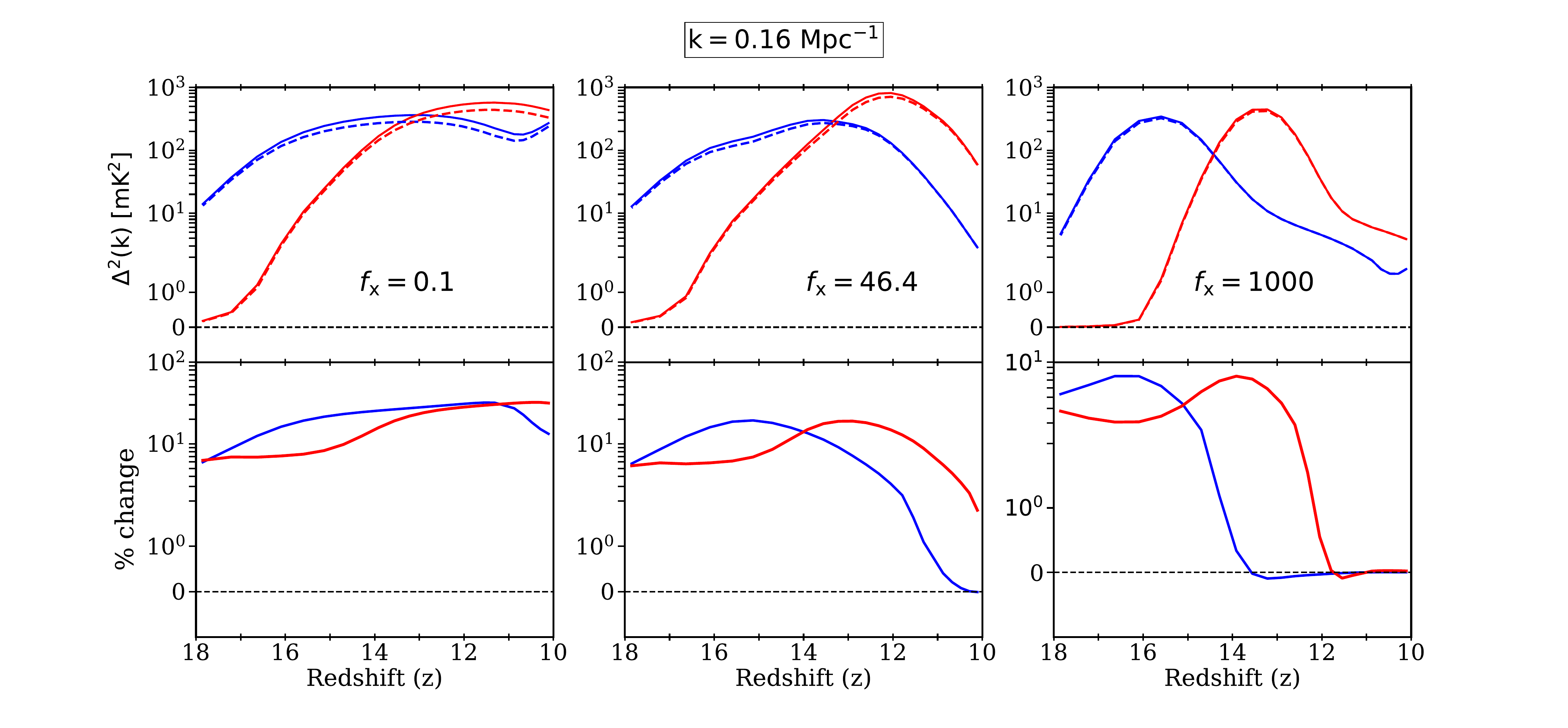}
\includegraphics[width=\textwidth]{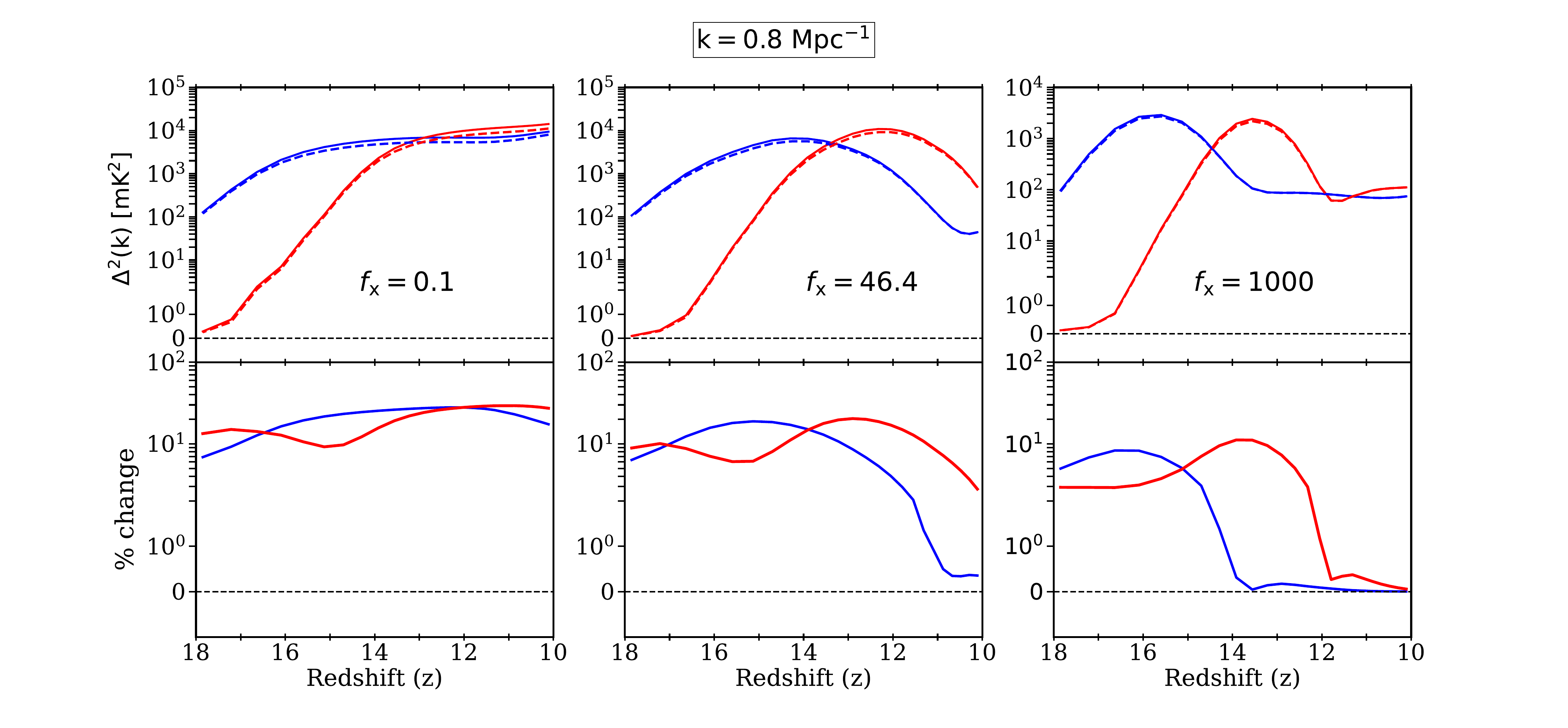}
    \caption{
    Redshift evolution of the spherically averaged dimensionless power spectrum for the approximated and exact $\delta T_b$, along with the percentage difference between them, at scales $\rm k = 0.16\ Mpc^{-1}$ (upper two rows) and $\rm k = 0.8\ Mpc^{-1}$ (lower two rows). In the first and third rows, the blue solid and dashed lines represent the power spectrum of the approximated and exact $\delta T_b$, respectively, for the $ \MHMIN = 10^9\ \MSUN$ model, whereas the red solid and dashed lines represent the same for the $\MHMIN = 10^{10}\ \MSUN$ model. In the second and fourth rows, the blue and red solid lines show the corrsponding percentage differences between the approximated and exact cases for the $\MHMIN = 10^9\ \MSUN$ and $\MHMIN = 10^{10}\ \MSUN$ models, respectively.}
    \label{fig:powspec}
\end{figure*}

Although our main goal is to investigate the impact of higher-order terms of $\tau$ on the non-Gaussian features of the 21-cm signal from the CD, it is also worthwhile to study the \HI power spectrum in this context. Measuring the power spectrum is the primary objective of ongoing and upcoming CD and EoR radio interferometric experiments. Significant progress has already been made in modelling the power spectrum \citep{Furlan2004, Mesinger2007, Datta2012a, Jensen2013}, placing upper limits using ongoing experiments \citep{2024MNRAS.534L..30A, 2019ApJ...883..133K,  2020MNRAS.493.4711T, Abdurashidova_2023}, and interpreting these measurements \citep{ 2020MNRAS.498.4178M, 2022ApJ...924...51A, 2020arXiv200603203G, 2025arXiv250500373G}.

In Fig. \ref{fig:powspec}, we show the redshift evolution of spherically averaged power spectra of both approximated and exact $\DTB$ at the length scales $k=0.16$ Mpc$^{-1}$ and $k=0.8$ Mpc$^{-1}$. We can see that, the power spectrum initially increases with decrease in redshift, and then decreases beyond a certain redshift.  It starts decreasing earlier for higher $\FX$ value. However, unlike bispectrum evolution which is a marker of different astrophysical events through its sign flip, power spectrum is positive throughout the entire CD period.

We can see, for $k=0.16$ Mpc$^{-1}$, that at the beginning of the CD power spectrum increases as the $\LYA$ radiation starts coupling the spin temperature with the kinetic temperature. However, as the average $\tau$ is small at the beginning of CD, the change in the power spectrum between approximated and exact cases is small. As time progresses, $\LYA$ coupling becomes a dominating astrophysical event and as $\LYA$ radiation spreads throughout the IGM, the average $\tau$ becomes maximum around that redshift. Therefore, the change in the power spectrum is also maximum at that redshift. After that, when X-ray radiation starts heating the IGM, the average $\tau$ starts decreasing, which also decreases the change in the power spectrum. Finally, when IGM gets almost entirely heated the power spectrum and its change both decreases.

Similar behaviour is seen in all other scenarios. Considering the model with $\FX=46.4$ and $\MHMIN=10^9\, \MSUN$ for $k_1 = 0.16\,{\rm Mpc}^{-1}$, we see the percentage change between approximated and exact power spectra  increases with decrease in redshift as average $\tau$ increases while $\LYA$ coupling dominates over X-ray heating. The  maximum change is $\approx 20\%$ around $z =  15$ for this model. 
For higher halo mass threshold, such as $\MHMIN=10^{10}\, \MSUN$ with same $\FX$ value we see similar changes but at later redshifts. For models with $\FX=0.1$, $\MHMIN=10^9\, \MSUN$ and $\FX=0.1$, $\MHMIN=10^{10}\, \MSUN$ the change goes up to $\approx 30\%$ at redshift $z\approx 11.5$ and $z\approx 10.5$, respectively. The change reaches up to merely $\approx 8\%$ at redshift $z\approx 16.5$ and $z\approx 14$ for models with $\FX=1000$, $\MHMIN=10^9\, \MSUN$ and $\FX=1000$, $\MHMIN=10^{10}\, \MSUN$, respectively. 

Similar qualitative behaviour is found at smaller length scales. The percentage changes in the power spectrum at $k=0.8$ Mpc$^{-1}$ are $\approx 28\%$ and $\approx 30\%$ for models with $\FX=0.1$, $\MHMIN=10^9\, \MSUN$ and $\FX=0.1$, $\MHMIN=10^{10}\, \MSUN$, respectively. For higher $\FX$ values, the changes are similar to that of $k=0.16 \, {\rm Mpc}^{-1}$. For example,  changes for $\FX=46.4$, $\MHMIN=10^9\, \MSUN$ and $\FX=46.4$, $\MHMIN=10^{10}\, \MSUN$ models are up to $\approx 20\%$ and for $\FX=1000$, $\MHMIN=10^9\, \MSUN$ and $\FX=1000$, $\MHMIN=10^{10}\, \MSUN$ models the changes are as high as $\approx 8\%$ and $\approx 11\%$ at $z\approx 16.5$ and $z\approx 13.5$, respectively.

\section{Importance of different Higher-Order Terms of $\tau$ in Bispectrum Estimation}
\label{sec:5}

\begin{figure*}[!t]
\includegraphics[width=0.45\textwidth]
{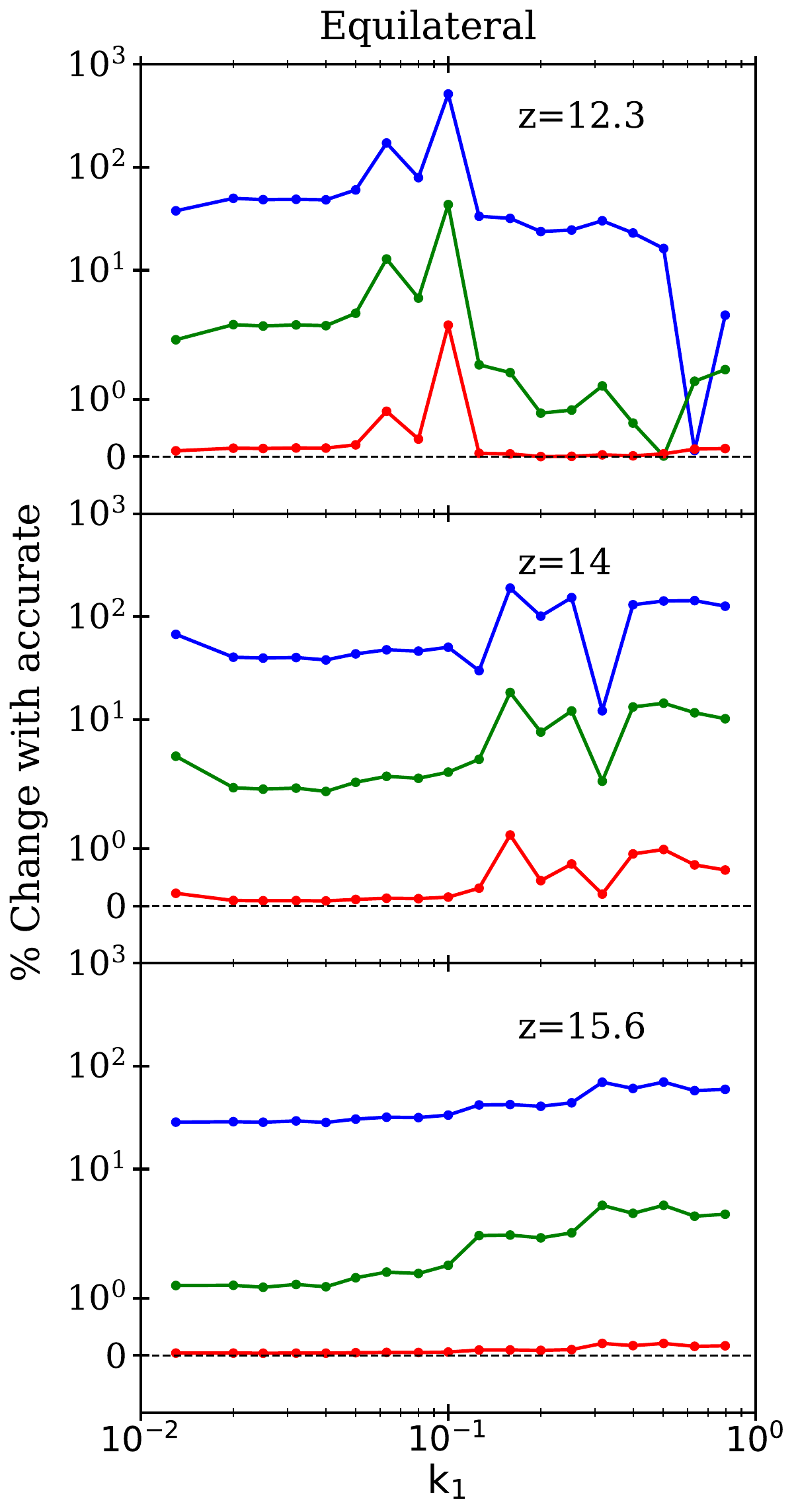}
\hspace{0.3cm}
\includegraphics[width=0.418\textwidth]{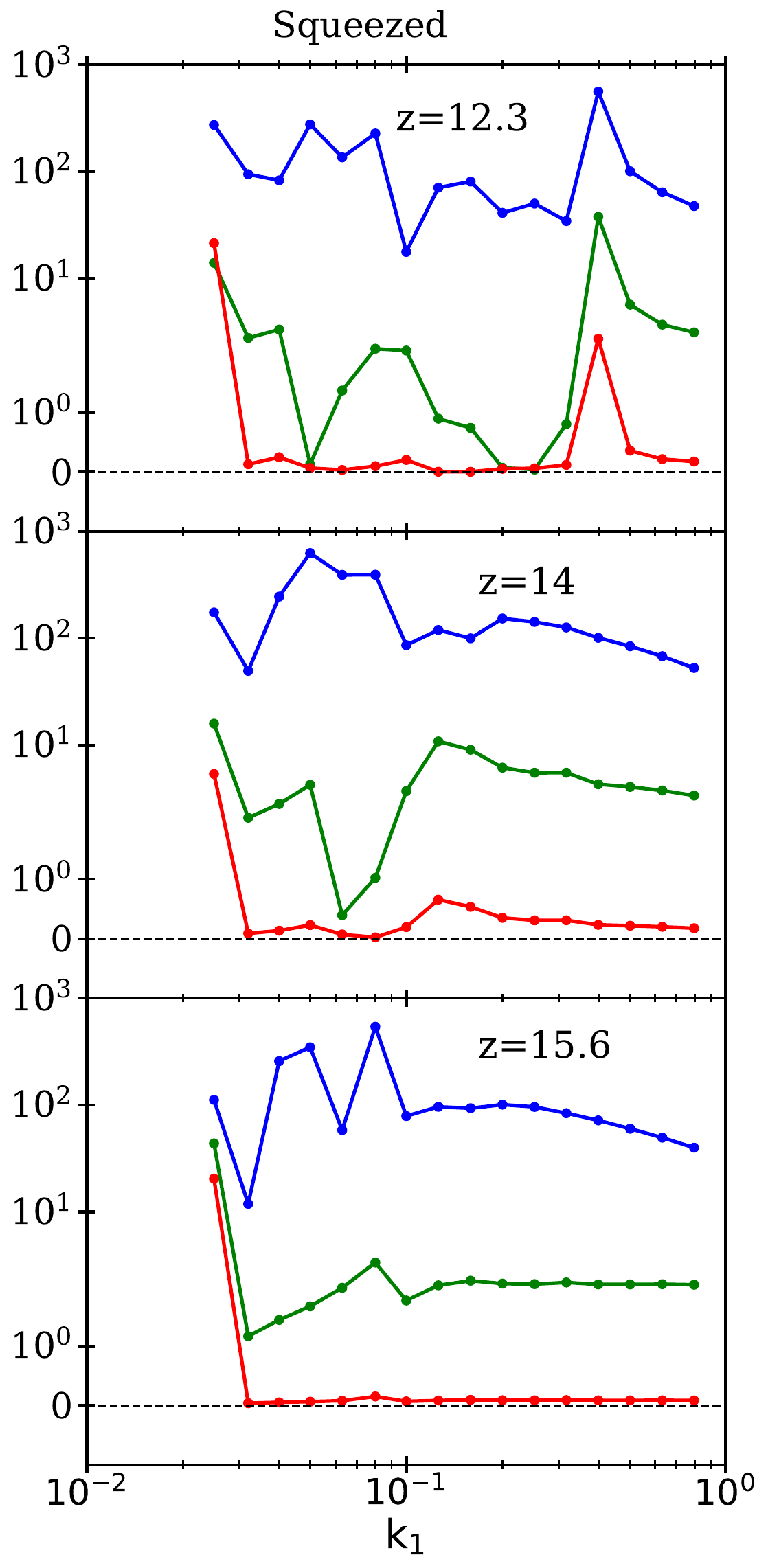}
\caption{
Percentage differences between the exact and approximated bispectra of $\DTB$ for the model with $\FX = 0.1$ and $\MHMIN= 10^9\ \MSUN$, at redshifts $z = 12.3$, $14$, and $15.6$. The approximated $\DTB$ includes contributions up to the first (blue), second (green), and third order (red) terms of $\tau$. The left and right panels correspond to the equilateral triangle configuration and the squeezed limit, respectively. }
\label{fig:third_order}
\end{figure*}

It would be interesting to investigate how many higher-order terms of $\tau$ need to be retained in equation (\ref{eq:3rd_ord}) to accurately model $\DTB$, especially for capturing the non-Gaussian features of the signal. Furthermore, it is important to understand how each higher-order term of $\tau$ contributes to the bispectrum. Therefore, expanding the exact expression of $\DTB$ (\ref{eq:exact}) in powers of $\tau$ and determining which orders are important would be a useful step for accurate modelling and analysis.

In Fig.~\ref{fig:third_order}, we show the percentage difference between the exact bispectrum and those estimated using up to the first, second, and third order terms of $(1 - e^{- \tau})$ for $\FX=0.1$ and $\MHMIN=10^9 \, \MSUN$.  We see that, the difference is maximum, reaching up to a few $100 \%$, when using only up to the first-order term for all scenarios considered. However, the difference reduces significantly to $\sim 10\%$ after including the second-order term. When terms up to the third order are included in estimating  $\DTB$, the differences become negligibly small, typically  $\sim 1\%$ or less. We have shown the result for three different redshifts $12.3$, $14$ and $15.6$. We choose this range of redshifts as average $\tau$ is relatively high in this range, leading to a large difference in the percentage change. However, this trend holds true for all values of $k_1$, redshifts, models, and triangle configurations considered in this study.

\section{Summary and Discussions}
\label{sec:6}
The \HI 21-cm optical depth, which is proportional to the neutral hydrogen density and inversely proportional to the spin temperature, can become quite large during the Cosmic Dawn (CD). This is because the spin temperature is expected to remain very low before X-ray heating begins. Our simulations show that, in some CD models, the optical depth can reach values as high as $\approx 0.55$, with nearly $50 \%$ of the simulation pixels showing optical depths greater than $0.1$. Therefore, the commonly used approximation that the optical depth is much less than one ( $\tau <<1 $) which simplifies the expression $[1- \exp(-\tau)] \approx \tau$ may not hold during this epoch.

To revisit the impact of this assumption on $\delta T_b$ and its associated statistics, we performed a suite of numerical simulations spanning a wide range of astrophysical parameters, including the X-ray heating efficiency ($\FX$) and the minimum halo mass ($\MHMIN$). These simulations model $\LYA$  coupling, X-ray heating, spin temperature evolution, and ionization using a 1D radiative transfer code, named {\sc grizzly}. We used them to study how large optical depths affect the statistical properties of the 21-cm signal, including the power spectrum, skewness, and bispectrum of the differential brightness temperature $\DTB$. 

We find that $\DTB$ is significantly suppressed when the full expression with  $[1- \exp(-\tau)]$ is used, particularly in regions with high neutral hydrogen density and low spin temperature. This suppression is not uniform across the volume but is stronger in regions where the ratio $\rho_{\rm HI}/T_s$ is higher. As a result, an additional non-Gaussian component arising from higher-order terms of $\tau$ emerges, with a sign opposite to that of the existing bispectrum. In the model with $\MHMIN=10^9 \, \MSUN$ and $\FX=0.1$, the skewness changes by up to $ \sim 500\%$. For the same halo mass, the change is about $40\%$ and $60 \%$ for $\FX=46.4$ and $\FX=1000$, respectively. For a higher halo mass threshold of $\MHMIN=10^{10} \, \MSUN$, the skewness changes by approximately $75 \%$, $25\%$, and $20\% $ for $\FX=0.1$, $46.4$, and $1000$, respectively.

We also studied the impact on the bispectrum in both equilateral and squeezed triangle configurations across all redshifts during CD. We analyzed two different scales, $k_1=0.16 \, {\rm Mpc}^{-1}$ and $0.8 \, {\rm Mpc}^{-1}$, representing large and small scales. For low-heating models ($\FX = 0.1$) with both $\MHMIN = 10^9, \MSUN$ and $10^{10}, \MSUN$, where the optical depth remains high throughout the CD, the change in the bispectrum is consistently large—reaching up to $\sim300\%$ at both scales and triangle configurations.  However, in models with moderate and high X-ray heating efficiency ( $\FX=46.4$ and $1000$), the changes are  smaller—typically around $\sim 10\%$ to $\sim 200\%$, regardless of halo mass or triangle shape.

Further, we investigated how many terms in the $\tau$-expansion are needed to accurately model  $\DTB$, especially to capture its non-Gaussian features. We find that including terms up to third order in $\tau$ reduces the difference to less than $\sim 1\%$. This result holds across all
$k$-values, redshifts, models, and triangle configurations considered in our study.

Our analysis suggests that for non-gaussian study of $\DTB$ signal we need to incorporate higher order terms of $\tau$ along with the linear order term. It demonstrates that higher order terms of $\tau$ can have considerable contribution to the non-Gaussian statistics during early stages of CD. However, we can still use the approximated form of $\DTB$ (eq.\ref{eq:approx}) towards the later stages of CD and during EoR when the spin temperature is high i.e. $\tau$ is low.

Finally, we note that our entire analysis has been carried out in real space. Including the effect of peculiar velocities would further enhance the optical depth. This could amplify the impact on the 21-cm signal.

\section*{Data Availability}
The data underlying this work will be shared upon reasonable request to the corresponding author.

\section{Acknowledgments}
IN acknowledges financial support from the Swami Vivekananda Merit-cum-Means Scholarship (SVMCM), Govt. of West Bengal, India. IN also acknowledges support from the DST-WISE Fellowship (DST/WISE-PhD/PM/2023/104), Govt. of India. KKD acknowledges financial support from ANRF
(Govt. of India) through a SERB  MATRICS scheme (MTR/2021/000384). The authors also acknowledge the use of computational resources provided by the Inter-University Centre for Astronomy and Astrophysics (IUCAA), Pune. AKS acknowledges support from National Science Foundation (grant no. 2206602). RG acknowledges support from SERB, DST Ramanujan Fellowship no. RJF/2022/000141. AM  acknowledges financial support from Council of Scientific and Industrial Research (CSIR) via  CSIR-SRF fellowships under grant no. 09/0096(13611)/2022-EMR-I. MK acknowledges financial support from the foundations Carl Tryggers stiftelse för vetenskaplig forskning (grant agreement no. CTS 21:1376).



\bibliographystyle{mnras}
\bibliography{ref} 







\bsp	
\label{lastpage}
\end{document}